\documentclass[10pt]{article}
\usepackage{fullpage}
\usepackage[utf8]{inputenc}

\usepackage[explicit]{titlesec}
\usepackage{hyphenat}
\usepackage{ragged2e}
\RaggedRight
\usepackage{amsfonts}
\usepackage{ebgaramond-maths}
\usepackage[cmintegrals,cmbraces]{newtxmath}

\usepackage{amsmath,amsthm}
\usepackage{graphicx}
\usepackage{url}
\usepackage{newtxtext,newtxmath}
\usepackage{fancyhdr}
\usepackage{geometry}
\usepackage[utf8]{inputenc}

\usepackage{amsmath}
\usepackage{natbib}

\RequirePackage[colorlinks,linkcolor=blue,citecolor=blue,urlcolor=blue]{hyperref}
\usepackage[utf8]{inputenc}
\usepackage{graphicx}
\usepackage[dvipsnames]{xcolor}

\newtheorem{definition}{Definition}

\usepackage{authblk}

\usepackage{natbib}
\def\ba{\boldsymbol{a}}

\def\bg{\boldsymbol{g}}

\def\bx{\boldsymbol{x}}
\def\by{\boldsymbol{y}}

\def\bD{\boldsymbol{D}}

\def\zeros{\boldsymbol{0}}

\def\bSigma{\boldsymbol{\Sigma}}

\title{A Comparison of Zero-Inflated Models for Modern Biomedical Data}
\author{\textit{Max Beveridge$^a$, Zach Goldstein$^b$, and Hee Cheol Chung$^{c}$}\\
\small
\textit{$^a$ Department of Statistics, The George Washington University, Washington, DC}\\
\textit{$^b$ Department of Mathematics, Yeshiva University, New York, NY}\\
\textit{$^c$ Department of Mathematics and Statistics, University of North Carolina at Charlotte, Charlotte, NC}\\
\medskip
\textit{Students: maxbeveridge03@gwu.edu, zgoldst3@mail.yu.edu}\\
\textit{Mentor: hchung13@uncc.edu
}}

\begin{document}

\maketitle

\section*{ABSTRACT}
Many data sets cannot be accurately described by standard probability distributions due to the excess number of zero values present. For example, zero-inflation is prevalent in microbiome data and single-cell RNA sequencing data, which serve as our real data examples. 
    Several models have been proposed to address zero-inflated datasets including the zero-inflated negative binomial, hurdle negative binomial model, and the truncated latent Gaussian copula model. This study aims to compare various models and determine which one performs optimally under different conditions using both simulation studies and real data analyses. We are particularly interested in investigating how dependence among the variables, level of zero-inflation or deflation, and variance of the data affects model selection.  

\section*{KEYWORDS}
Zero-Inflated Models; Hurdle Models; Truncated Latent Gaussian Copula Model; Microbiome Data; Gene-Sequencing Data; Zero-Inflation, Negative Binomial; Zero-Deflation

\section*{INTRODUCTION}

Zero-inflated data refers to datasets with an excess of zeros, where the proportion of zeros cannot be adequately captured by standard probability distributions. Such data frequently arise in various fields, such as health and epidemiology, where large numbers of zeros are often encountered. For example, in substance abuse research, the majority of individuals do not engage in substance abuse, leading to a predominance of zero observations \citep{feng2021comparison}. Similarly, zero-inflated data is common in biomedical research including microbiome studies and single-cell RNA sequencing, where zeros occur due to limited sequencing depth \citep{vandeputte2017quantitative, li2015microbiome}. Given the widespread occurrence of zero-inflated data across numerous disciplines, it is essential to model these datasets accurately to ensure valid analyses. Failure to properly account for zero inflation can lead to poor estimation and the potential oversight of statistically significant findings. Accurate modeling of zero-inflated data not only improves the estimation of key parameters but also reduces bias and enhances the understanding of dependence structures \citep{Peruman-Chaney2013}. Violating distributional assumptions of statistical tests is one of the ``seven deadly sins'' of comparative analysis and could result in biased or incorrect parameter estimates and incorrect $p$-values \citep{https://doi.org/10.1111/j.1420-9101.2009.01757.x}. With regards to zero-inflated data, several studies have found that misspecifying the distribution of a general linear model (GLM) when data is zero-inflated leads to invalid statistical inference (e.g. using a Poisson or Negative Binomial (NB) regression model when the data follows a zero-inflated Poisson or zero-inflated NB distribution) \citep{https://doi.org/10.1111/2041-210X.13559}.

Zero-inflated models, including zero-inflated Poisson (ZIP), zero-inflated negative binomial (ZINB), hurdle Poisson (HP), and hurdle negative binomial (HNB), have been widely used to model zero-inflated data across fields such as ecology, environmental science (e.g., species counts), economics (e.g., consumer purchases), insurance (e.g., claims data), and criminology (e.g., crime counts in different areas).  The key difference between zero-inflated and hurdle models lies in how they handle the excess number of zeros. Zero-inflated models combine a point mass at zero with a standard distribution that also allows non-zero probability at zero. The point mass accounts for structural zeros (inherent zeros), while the non-zero probability from the standard distribution models sampling zeros (zeros that occur by chance). In contrast, hurdle models only account for structural zeros by using a mixture of a point mass at zero and a standard distribution that is truncated above zero.

Several studies have been published comparing the applicability of these models. ZIP and ZINB models were useful when modeling certain psychological outcomes; however, faced limitations with intolerance to model misspecification and overdispersion (when the variance of the data is greater than what the model predicts) \citep{hua2014structural}. When comparing standard Poisson and NB models to ZIP, ZINB, HP, and HNB models in adverse vaccine side-effect data, ZINB and HNB models performed best as they could better model variance since they don't require the mean and variance to be the same as do the Poisson models \citep{rose2006use}. They concluded the choice between a zero-inflated and a hurdle model was dependent on the presence of structural and sampling zeros. In addition, we want to point out that, with absence of covariates, these zero inflation and hurdle models cannot capture dependence between variables. 

Nevertheless, a major limitation of zero-inflated and hurdle models is their inability to adequately account for dependencies between variables when dealing with multivariate data. While the presence of covariates or auxiliary variables allows for dependence modeling within the generalized linear model framework, such covariates are often unavailable in real-world applications. A latent variable modeling approach \citep{dong2014multivariate}, which assumes each variable is a sum of unique and shared negative binomial latent variables, can be used. However, it does not address the limitation of zero-inflation models' sensitivity to overdispersion. More recently, the truncated latent Gaussian copula (TLNPN) model \citep{fan2017high,yoon2020sparse} was introduced to address the large number of zeros and extreme skewness commonly found in biomedical data, while also properly modeling the dependence between variables \citep{yoon2020sparse}.  The TLNPN models are specifically designed for high-dimensional, mixed data types to uncover linear relationships between variables. By leveraging the Gaussian copula, the TLNPN model treats variables as latently Gaussian and uses Kendall's $\tau$ to estimate latent correlations between them.


With the recent emergence of the TLNPN model, there has been a lack of research on its relative performance compared to the other models. This study seeks to fill that gap by investigating what characteristics of data decide the best model. In particular, we want to investigate how zero-inflation and deflation, the level of dependence among the variables, and the variance of the data impacted model choice. We seek to evaluate the advantages and disadvantages for models that are commonly used to analyze zero-inflated data.

\section*{Method and Procedures}

\subsection*{\textit{Models for Zero-Inflated Data}}
\begin{figure}
    \centering
    \textbf{Negative Binomial vs. Zero-Inflated Negative Binomial Distribution}\par\medskip
    
    \includegraphics[width=0.49\linewidth]{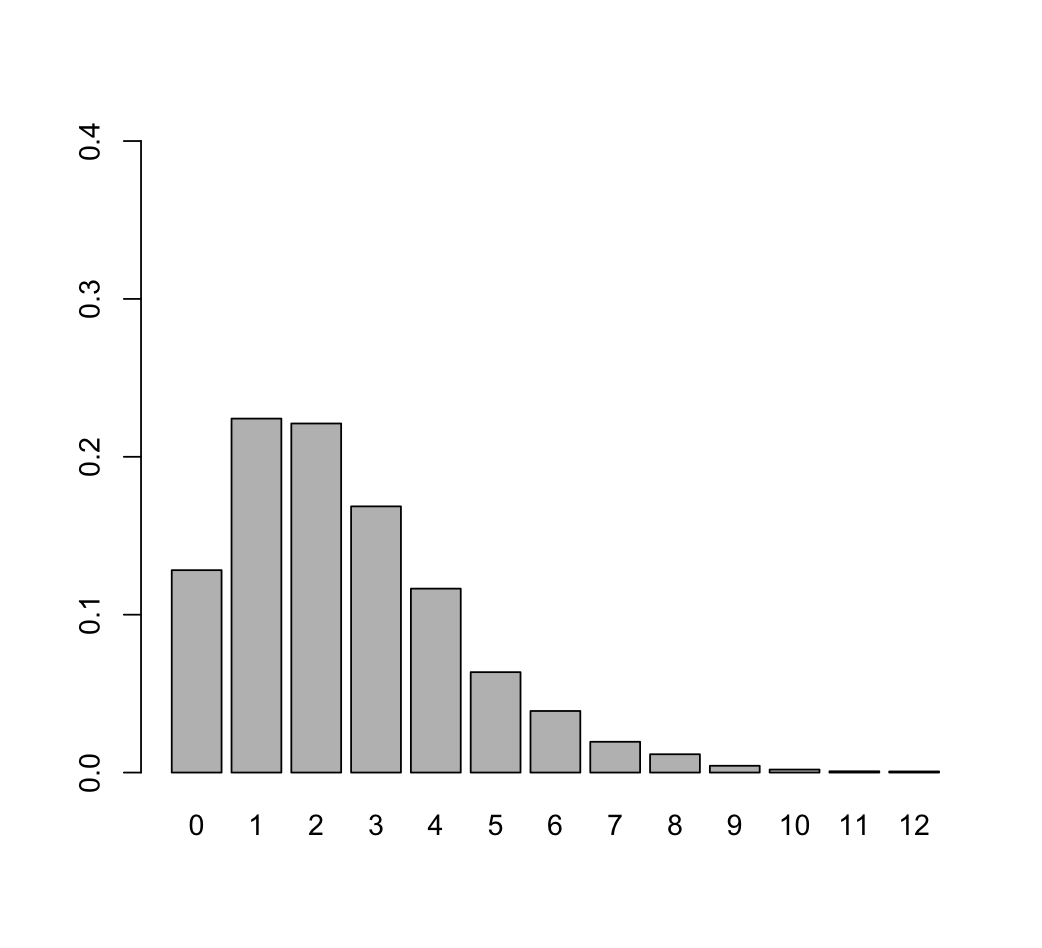}\includegraphics[width=0.49\linewidth]{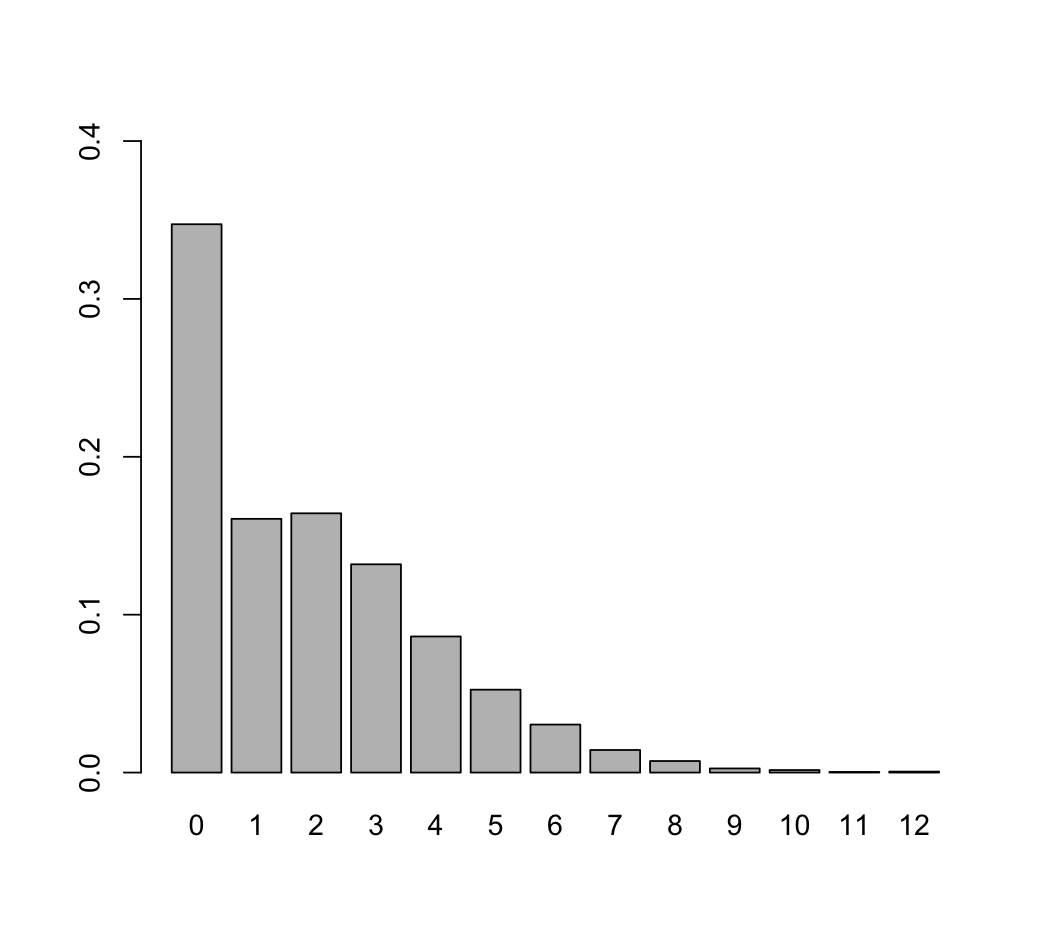}
    \caption{Shown is a standard negative binomial distribution with $\mu = 2.5$, $r=5$, and the probability that $Y=0$ is $0.1317$ (left) and zero-inflated negative binomial distribution with $\mu = 2.5$, $r = 5$, and $\pi_Z = 0.25$, and the probability that $Y=0$ is $0.3487$ (right).}
    \label{ZINB}
\end{figure}

Zero-inflated models account for an excess number of zeros by adjusting the probability of observing zero of a standard probability distribution. In particular, the form of the probability mass function (pmf) of a zero-inflated model is given by:
\begin{equation*}
P(Y = y) = 
\begin{cases}
\pi_Z + (1-\pi_Z)p(y = 0; \mu) &\text{for $y=0$},\\
(1-\pi_Z)p(y;\mu)&\text{if $y>0$},
\end{cases}
\end{equation*}
where $p(\cdot)$ is a pmf of a discrete random variable following a standard distribution, e.g., Poisson or negative binomial distribution, $\mu$ is the mean of the distribution, and $\pi_Z$ is the weight parameter controlling the degree of zero inflation. 
One of the popularly used zero-inflated models is the zero-inflated Negative Binomial (ZINB) model with pmf:
\begin{equation*}
P(Y = y) = 
\begin{cases}
\pi_Z + (1-\pi_Z)(\frac{r}{\mu + r})^r &\text{for $y=0$},\\
(1-\pi_Z)\frac{\Gamma(y+r)}{\Gamma(r)y!}(\frac{\mu}{\mu+r})^{y} (\frac{r}{\mu + r})^r&\text{if $y>0$},\\
\end{cases}
\end{equation*}
where $\mu$ is the mean of the Negative Binomial model, $r$ is the dispersion parameter, and $\pi_Z$ is the probability of extra zeros. In zero-inflated models, the probability of observing a zero is given by $\pi_Z + (1 - \pi_Z)p(y=0;\mu)$. As a result, the probability is bounded below by $p(y=0;\mu)$, which corresponds to the probability under the standard negative binomial model. Consequently, zero-inflated models cannot account for zero deflation. An illustrative example of the zero-inflated Negative Binomial distribution is provided in \textbf{Figure \ref{ZINB}}. 
In contrast to zero-inflated models, huddle models are able to account for zero-inflation and zero-deflation. 

\begin{figure}
    \centering
    \textbf{Hurdle Negative Binomial Distributions}\par\medskip
    \includegraphics[width=0.49\linewidth]{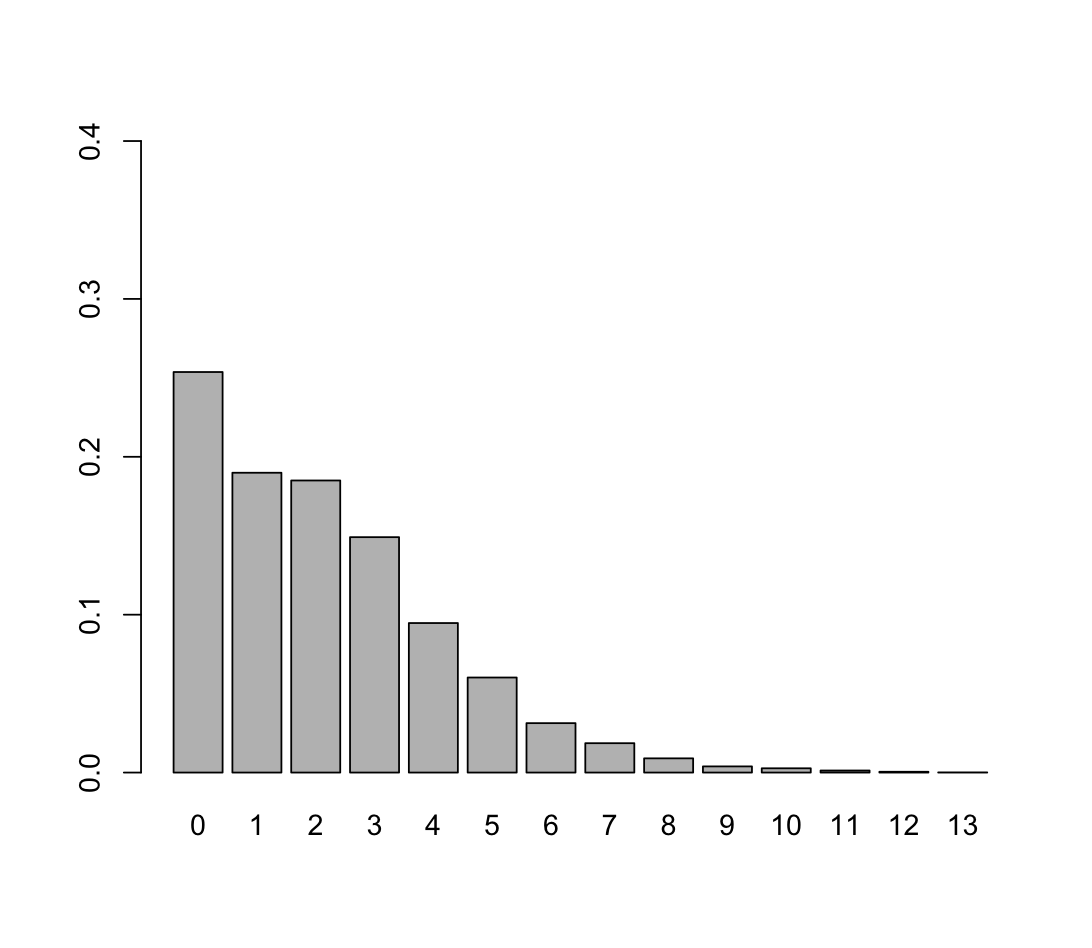}\includegraphics[width=0.49\linewidth]{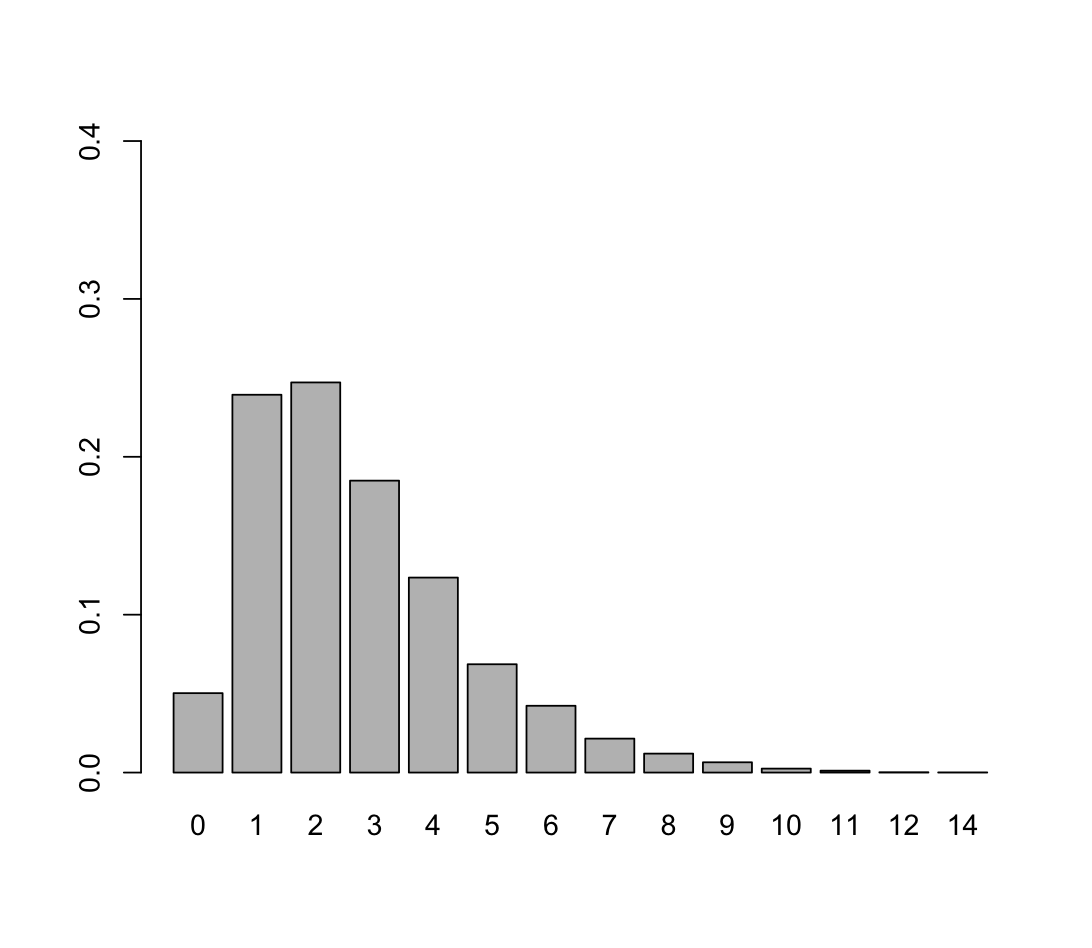}
    \caption{Shown are hurdle negative binomial distributions with $\mu = 2.5$ and $r = 5$, where $P = 0.25$, so the probability $Y=0$ is $0.25$ (left) and $P = 0.05$, so  the probability that $Y=0$ is $0.05$ (right). Under a standard negative binomial distribution, the probability that $Y=0$ is $0.1317$.}
    \label{HNB}
\end{figure}

Hurdle models are distinct from zero-inflated models because they only account for structural zeros and they are able to model zero-deflation as opposed to zero-inflated models. Zero-deflation occurs when there are less zero values present than a standard probability distribution would predict. The form of the pmf of a hurdle model is:
\begin{equation*}
P(Y = y) = 
\begin{cases}
\pi_H &\text{for $y=0$}\\
(1-\pi_H)\frac{p(y;\mu)}{1-p(y = 0;\mu)}&\text{if $y>0$}\\
\end{cases}
\end{equation*}
where $p(\cdot;\mu)$ is the pmf of a poisson or negative binomial distribution with mean $\mu$. The parameter $\pi_H$ is the probability that a structural zero occurs and can take any value from $0$ to $1$. 
The hurdle negative binomial (HNB) model is given by:
\begin{equation*}
P(Y = y) = 
\begin{cases}
\pi_H &\text{for $y=0$}\\
\frac{1-\pi_H}{1-(\frac{r}{\mu+r})^r}
\frac{\Gamma(y +r)}{\Gamma(r)y!}
(\frac{\mu}{\mu+r})^{y}
(\frac{r}{\mu+r})^r
&\text{if $y>0$}.\\
\end{cases}
\end{equation*}
Under the hurdle model, zero occurs with probability $\pi_H$, which can be smaller than the probability of $Y=0$ under the negative binomial model and thus capable of modeling zero deflated variables. Examples of the hurdle negative binomial distribution can be seen in \textbf{Figure \ref{HNB}}.
When data involves multiple zero-inflated variables, their associations can be modeled within the generalized linear model framework, assuming covariates are available.

For multiple zero-inflated random variables $Y_1,\ldots,Y_p$, given the covariates $\boldsymbol{x} = (x_1,\ldots,x_{q_1})^{\top}$ and $\boldsymbol{z} = (z_1,\ldots,z_{q_2})^{\top}$, which are shared across $Y_1,\ldots,Y_p$, their associations can be modeled within the generalized linear model (GLM) framework. In particular, the ZINB regression model is given by
\begin{flalign}\label{eq:log_logit_ZINB}
\ln(\mu_{ij}) = \boldsymbol{x}_i^T \boldsymbol{\beta}_j , \text{ and } 
\text{logit}(\pi_{Z,ij}) = \boldsymbol{z}_i^T\boldsymbol{\gamma}_j
\end{flalign}
where $\boldsymbol{\beta}_j \in \mathbb{R}^{q_1}$ and $\boldsymbol{\gamma}_j \in \mathbb{R}^{q_2}$ are the regression coefficients for the mean $\boldsymbol{\mu}_{i}=(\mu_{i1},\ldots,\mu_{ip})^{\top}$ and $\boldsymbol{\pi}_{Z,i} = (\pi_{Z,i1},\ldots, \pi_{Z,ip})^{\top}$, respectively, and $\text{logit}(\pi_Z) = \log\{ \pi_Z /(1-\pi_Z) \}$. For each $j=1,\ldots,p$, the parameters $\boldsymbol{\beta}_j$ and $\boldsymbol{\gamma}_j$, and the dispersion parameter $r_j$, can be estimated using the maximum likelihood estimator, with the log-likelihood function $L_{ZI}$ defined as $L_{ZI} = L_1 + L_2 + L_3 - L_4$, where
\begin{flalign*}
L_1  &= \sum_{i:y_i=0} \ln\bigg\{ e^{\boldsymbol{z_i}^T\boldsymbol{\gamma}_j} + \left( 1 + \frac{\mu_{ij}}{r_j}\right)^{-r_j}\bigg\}, \quad
L_2  = \sum_{i:y_{ij}>0}\sum^{y_{ij}-1}_{t=0} \ln(t+r_j)\\
L_3  &= \sum_{i:y_i>0}\big\{-\text{ln}(y_{ij}!) - (y_{ij} + r_j)\ln\left(1+\frac{\mu_{ij}}{r_j}\right)+y_{ij}\ln(r_j^{-1})+y_{ij}\ln(\mu_{ij}) \big\} \\
L_4 &= \sum^{n}_{i=1} \ln(1+e^{\boldsymbol{z_i}^T\boldsymbol{\gamma}_j} ).
\end{flalign*}




The hurdle negative binomial regression model is given by
\begin{flalign}\label{eq:log_logit_HNB}
\log(\mu_{ij}) = \boldsymbol{x}_{i}^T \boldsymbol{\beta}_j , \text{ and } \text{logit}(\pi_{H,ij}) = \boldsymbol{x}_{i}^T\boldsymbol{\gamma}_{j}
\end{flalign}
The log-likelihood function, $L_{\text{H}}$, is given as
\begin{equation*}
L_{H} = \sum^{n}_{i = 1} (I_{y_{ij}=0} \ln(\pi_{H,ij}) + I_{y_i>0}(\ln(1-\pi_{H,ij}) + \ln(h(y_{ij};\mu_{ij}, r_j) - \ln(1-(1+r_j\mu_{ij})^{-r_j}))
\end{equation*}
where $h(y_{ij};\mu_{ij}, r_j)$ denotes the pmf of negative binomial distribution with mean $\mu_{ij}$ and dispersion parameter $r_j$ \citep{ehsan2012hurdle}.

Nevertheless, in real applications, such covariates are often not readily available. In these cases, we can only fit the intercept parameters $\beta_0$ and $\gamma_0$, assuming that all variables are mutually independent. The Gaussian copula model addresses this limitation by utilizing a rank-based correlation estimator.
The Gaussian copula model assumes that, for a random vector $\boldsymbol{y} = (Y_1,...,Y_p)^{\top}$, there exist strictly increasing functions, $g_1,\ldots,g_p$, such that $\boldsymbol{z} = (g_1(Y_1),...,g_p(Y_p))^{\top}  \sim N_p(\boldsymbol{\mu}, \boldsymbol{\Sigma})$ for some mean $\boldsymbol{\mu}$ and covariance matrix $\boldsymbol{\Sigma}$.
It is important to note that $\boldsymbol{\mu}$ and $\boldsymbol{\Sigma}$ are not identifiable because, for any constants $a_j$ and $b_j$, the Gaussian copula model still holds with $g_j^* = a_j + b_j g_j$, $j=1,\ldots,p$, i.e., $(g^{*}_1(Y_1),...,g^{*}_p(Y_p))^{\top}$ follows $N_p(\boldsymbol{a} + \boldsymbol{\mu}, \boldsymbol{B}\boldsymbol{\Sigma}\boldsymbol{B})$, where $\boldsymbol{a} = (a_1,\ldots,a_p)^{\top}$ and $\boldsymbol{B}=\text{diag}\{b_j\}_{j=1}^{p}$. The identifiability issue is commonly addressed by assuming that $\boldsymbol{\mu} = \zeros_p$ and $\boldsymbol{\Sigma}$ is a positive definite correlation matrix \citep{liu2009nonparanormal,fan2017high}. If $g_j$s are differentiable, then we have analytic expression as $g_j =  \Phi^{-1}\circ F_j^{-1}$, where $F_{j}$ and $\Phi^{-1}$ are the distribution functions of $Y_j$ and standard Gaussian.
The Gaussian copula models are often denoted as $\boldsymbol{y} \sim NPN(\zeros_p,\bSigma,\bg)$ \citep{chung2022sparse}.

The Gaussian copula models assume that $Y_j$ are continuous and thus not valid for zero-inflated variables. To accommodate zero-inflated and highly skewed variables, the truncated Gaussian copula models  \citep{yoon2020sparse} have been introduced by incorporating an additional truncation mechanism, as follows:
\theoremstyle{definition}
\begin{definition}[Truncated Latent Gaussian Copula Model]
A random vector $\by \in \mathbb{R}^p$ satisfies the truncated latent Gaussian Copula model if there exists a random vector $\by^*\sim NPN(\zeros_p,\bSigma,\bg)$ and constants $D_j$, $j = 1,...,p$ such that $Y_j = I(Y_j^* > D_j)Y_j^*$ where $I(\cdot)$ is an indicator function. We then denote $\boldsymbol{y} \sim TLNPN(\zeros,\bSigma,\bg,\bD)$.
\end{definition}

The latent correlation matrix $\boldsymbol{\Sigma}$ of $TLNPN(\zeros,\bSigma,\bg,\bD)$ is estimated using Kendall's $\tau$. The sample Kendall's $\tau$ between the $j$th and $k$th variables is defined as:
\begin{equation*}
\hat{\tau}_{jk} = \frac{2}{n(n-1)} \sum_{1\le i\le i' \le n} \text{sign}(Y_{ij}-Y_{i'j})\text{sign}(Y_{ik}-Y_{i'k}).
\end{equation*}
There exists \citep{fan2017high,yoon2020sparse} an increasing bridge function $G$ defined so $G(\Sigma_{jk}) = E(\hat{\tau}_{jk}) = \tau_{jk}$ where $\Sigma_{jk}$ is an element of $\boldsymbol{\Sigma}$ corresponding to variables $Y_j$ and $Y_k$. The bridge function $G$ for two truncated variables is defined as:
\begin{equation*}
G_{TT}(\Sigma_{jk}; \Delta_j, \Delta_k) = -2\Phi_4(-\Delta_j, -\Delta_k, 0, 0; \boldsymbol{\Sigma}_{4a}) + 2\Phi_4(-\Delta_j, -\Delta_k, 0, 0; \boldsymbol{\Sigma}_{4b}),
\end{equation*}
where $\Delta_j = f_j(D_j)$ 
and $\Phi_4(a_1,a_2,a_3,a_4; \bSigma_4)$ denotes the cdf of $4$-dimensional Gaussian with zero mean and correlation matrix $\bSigma_4$ evaluated at $\ba=(a_1,a_2,a_3,a_4)^{\top}$.
The correlation matrices $\bSigma_{4a}$ and $\bSigma_{4b}$ are given by
\begin{equation*}
\boldsymbol{\Sigma}_{4a} = \begin{pmatrix}
1 & 0 & 1/\sqrt{2} & \Sigma_{jk}/\sqrt{2} \\
0 & 1 & \Sigma_{jk}/\sqrt{2} & 1/\sqrt{2} \\
1/\sqrt{2} & \Sigma_{jk}/\sqrt{2} & 1 & - \Sigma_{jk} \\
-\Sigma_{jk}/\sqrt{2} & 1/\sqrt{2} & -\Sigma_{jk} & 1
\end{pmatrix},
\end{equation*}
\begin{equation*}
\boldsymbol{\Sigma}_{4b} = \begin{pmatrix}
1 & \Sigma_{jk} & 1/\sqrt{2} & \Sigma_{jk}/\sqrt{2} \\
\Sigma_{jk} & 1 & \Sigma_{jk}/\sqrt{2} & 1/\sqrt{2} \\
\Sigma_{jk}/\sqrt{2} & 1/\sqrt{2} & 1 & \Sigma_{jk} \\
1/\sqrt{2} & \Sigma_{jk}/\sqrt{2} & \Sigma_{jk} & 1
\end{pmatrix}.
\end{equation*}
Using the bridge function $G$, we can consistently \citep{fan2017high,yoon2020sparse} estimate the latent correlation matrix as $\hat{\Sigma}_{jk} = G^{-1}(\hat{\tau}_{jk}; \hat \Delta_j, \hat \Delta_k)$, where $\hat \Delta_j$ is the moment estimator as $\hat \Delta_j = \Phi( \hat\pi_j)$ and $\hat\pi_j = n^{-1}  \sum_{i=1}^{n} 1(Y_{ij}=0)$ is the sample proportion of zeros of the $j$th variable.

The latent Gaussian copula model for binary type data was first introduced in 2017 to model dependence among discrete Arabidopsis gene data \citep{fan2017high}. The TLNPN model was introduced in 2020 which also created the rank-based estimators for the latent correlation matrix and found it useful for modeling gene-expression and micro-RNA data \citep{yoon2020sparse}. The TLNPN model has shown to be useful when performing discriminant analysis for microbiome data due to its ability to model dependence among zero-inflated variables \citep{chung2022sparse}. At the same time, zero-inflated and hurdle models are also popularly used to model zero-inflated data. However, a lack of research has been done comparing the TLNPN model to the other zero-inflated models and investigating the characteristics of data in which the TLNPN model performs better than the other models. 

\subsection*{\textit{Procedure of Simulation Studies}}

We examine the performance and robustness of ZINB, HNB, and TLNPN models using synthetic datasets across various conditions. We simulate data from each of the three populations—ZINB, HNB, and TLNPN—and, in settings two and three, evaluate performance by calculating the Wasserstein distance between test data independently generated from an assumed population and data generated from the corresponding fitted model. The Wasserstein distance measures the distance between two probability distributions, where at population level, the Wasserstein distance is defined as 
\begin{equation*}
{\displaystyle W_{p}(\mu ,\nu )=\inf _{\gamma \in \Gamma (\mu ,\nu )}\left\{\mathbb {E} _{(x,y)\sim \gamma }d(x,y)^{p}\right\}^{1/p}}.   
\end{equation*}
At the sample level, the Wasserstein distance between multivariate data set $Y$ and $\hat Y$ is defined as 
\begin{equation*}
W_{p}(Y,\hat Y) = \inf_{\theta \in \mathcal{S}_n } \left( \frac{1}{n} \sum_{i=1}^{n} \| Y_{i} - \hat Y_{\theta(i)}\|^{p} \right)^{1/p},
\end{equation*}
where $\theta $ is a permutation in the symmetric group $\mathcal{S}_n $. Wasserstein distance is a metric used for goodness-of-fit and statistical inference between two probability distributions where a lower distance implies a better fit \citep{PanaretosandZemel2018}.
As a summary measure of our results, we use arithmetic mean change (AMC). We use AMC because we want to measure the real relative difference between the performance of the models under varying scales of data. Furthermore, AMC is a symmetric measure of relative change, so a 10\% improvement and 10\% decline in performance from the HNB model to the TLNPN model are both captured by AMC, for example. Let $\omega_{\text{HNB}} = W_{p}(Y,\hat Y_{\text{HNB}})$ and $\omega_{\text{TLNPN}} = W_{p}(Y,\hat Y_{\text{TLNPN}})$ where $\hat Y_{\text{TLNPN}}$ is a simulated multivariate data set generated from the TLNPN model, $\hat Y_{\text{HNB}}$ is a simulated multivariate data set generated from the HNB model, and $Y$ is a multivariate dataset from generated from the true model. The AMC of the Wasserstein distance generated between the HNB model and the true model to the Wasserstein distance generated between the TLNPN model and the true model is given by:
\begin{equation}
    \text{AMC} = \frac{\omega_{\text{TLNPN}} - \omega_{\text{HNB}}}{(\omega_{\text{TLNPN}} + \omega_{\text{HNB}})/2}.
\label{eq_AMC}
\end{equation}
Therefore, a positive AMC implies that the HNB model performs better, a negative AMC implies the TLNPN model performs better, and AMC=0 implies that the models perform the same. 

We explore performance of ZINB, HNB, and TLNPN under three simulation settings. In the setting one, we aim to validate previous findings on the robustness of univariate ZINB and HNB models to model misspecification, focusing specifically on varying levels of zero inflation or deflation and differing proportions of sampling and structural zeros. In this simulation setting, we compare model performance using AIC since they are both discrete distributions and we seek to replicate previous studies \citep{feng2021comparison}. 
In simulation setting two, we compare the HNB model to the TLNPN model under HNB population data. We seek to evaluate whether the proportion of zeros or the dependence among the variables has an impact on the relative performance of each model. 
In simulation setting three, we again compare the HNB and TLNPN models except with TLNPN population. 
We again seek to understand how zero-proportion and dependence among variables affects relative model performance. 

In the multivariate settings (settings two and three), we apply the following autoregressive (AR) and geometrically decaying eigenvalues (GD) correlation structures to induce dependence between zero-inflated variables. The AR correlation structure is given by:
\begin{equation}
\boldsymbol{\Sigma} = [\rho^{|j-j'|}]_{1\le j,j' \le p}.
\label{eq_AR}
\end{equation}
The covariance matrix of the GD structure is given by $\boldsymbol{\Sigma} = \boldsymbol{\Gamma N \Gamma}^T$ where $\boldsymbol{N}=\text{diag}\{\nu_j\}_{j=1}^{p}$ is a diagonal matrix with geometrically decaying eigenvalues defined by:
\begin{equation}
    \nu_j = \frac{5(\rho^{j-1}-\rho^j)}{1-\rho^p}, \quad j = 1,...,p,
    \label{eq_GD}
\end{equation}
where a lower value of $\rho$ leads to higher correlations (in the absolute value sense) between the covariates, and $\boldsymbol{\Gamma}$ is uniformly generated from the orthogonal group of order $p$. The simulation settings are detailed as follows.

\subsection*{\textit{Setting One: Comparing the univariate ZINB and HNB Models}}

In this setting, we simulated $n = 500$ data points using covariate $X_i$ where $X_i \sim N(0,1)$, $i=1,\ldots,500$, $\ln(\mu_i) = \beta_0 + \beta_1 x_i$, $\text{logit}(\pi_{Z,i} )= \gamma_0 + \gamma_1 x_i$ and $\text{logit}(\pi_{H,i} )= \gamma_0 + \gamma_1 x_i$ as in \textbf{Equation \ref{eq:log_logit_ZINB}} and \textbf{Equation \ref{eq:log_logit_HNB}}.
We performed simulations under three parameter conditions controlling for $\beta_0=\ln(12), \beta_1 = 2, \gamma_1=2$, and $r=0.5$ at 20\% zero-proportion, 40\% zero-proportion, and 60\% zero-proportion, which we adjusted using the $\gamma_0$ parameter under both the ZINB model and HNB model. We then fit each model to the simulated data and compared the model fit through AIC. 

To further investigate the impact that zero-deflation had on relative model performance, we conducted a follow-up simulation study. We simulated $n=700$ data points using covariate $X_i \sim N(0,1)$, $i = 1,\ldots, 700$, $\ln(\mu_i) = \beta_0 + \beta_1 x_i$, and $\text{logit}(\pi_{H,i} )= \gamma_0 + \gamma_1 x_i$ where $\beta_0 = \ln(6/7)$, $\beta_1 = 0.1$, $\gamma_1 = 0$, and $r=2$ \textbf{Equation \ref{eq:log_logit_HNB}}. We varied $\pi_{H,i}$ from $0.08$ to $0.7$ by adjusting the $\gamma_0$ parameter. 
In this case, under the standard negative binomial distribution, the probability of $Y=0$ is 0.5. For each iteration, corresponding to a different proportion of zeros, we fit the data using both HNB and ZINB models and compared their AIC values.

\subsection*{\textit{Setting Two: Comparing the HNB and TLNPN Models (with HNB Population Data)}}

This setting aims to empirically compare the goodness-of-fit of the HNB and TLNPN models when data are generated from the HNB model. We don't consider the ZINB model because it cannot model zero-deflation, which is one of the conditions that we investigate. We also consider the HNB model fitted with and without covariates; however, in most biomedical datasets, covariates are unavailable. We set $n=1200$ and $p=5$ and generate covariates $\bx_i$, $i=1,\ldots,n$, independently and identically from the multivariate Gaussian with zero mean and covariance matrix given in \textbf{Equation \ref{eq_AR}} and \textbf{Equation \ref{eq_GD}}. We then set 
$\ln(\mu_{ij}) = \beta_0 + \beta_1 x_{ij}$ and $\text{logit}({\pi_{H}}_{ij} )= \gamma_0 + \gamma_1 x_{ij}$, $i=1,\dots,n$ and $j=1,\ldots,p$ as in \textbf{Equation \ref{eq:log_logit_HNB}}. We set $\beta_0 = 2.75$ and $r = 6$ for all the simulations. The parameter $\beta_1$ controls the impact that the covariate has on $\mu_{ij}$ and thus, affects the scale and dependence among the variables. It was set at $0$, $1$, and $2$. The parameter $\gamma_0$ controls the zero-proportion of the variable and was set at $\text{ln}(1/20)$, $\text{ln}(1/9),$ and $ \text{ln}(1/3)$. We're interested in whether zero-inflation or zero-deflation impacts relative model performance. We also varied the parameter $\gamma_1$, which controls how the covariate affects the probability of a structural zero, which impacts the variance of the data. This parameter was set at $-0.8$, $0$, and $0.8$. Finally, we also investigated the impact of the parameter $\rho$, which controls the amount of correlation between the covariates under both the AR and GD correlation structures, and thus, it affects the dependence among the HNB variables. The parameter $\rho$ was set at $0.01$, $0.3$, $0.7$, and $0.9$. We wanted to repeat this simulation study with a GD correlation structure because it can create negative correlation between covariates, unlike the AR correlation structure, and it more closely resembles correlation matrices found in real life.

As a goodness-of-fit measure, we consider 5-fold cross-validated prediction error. In particular, we randomly split $n=1200$ observations into five equal-sized folds, using four of the folds for training and keeping the remaining one for testing. The HNB and TLNPN models are fitted to the training data, and from each fitted model, we simulate a dataset of 240 observations and obtain the Wasserstein distance between the data from the fitted model and the test data. The described process is repeated for each fold, and we average the resulting five Wasserstein distances to obtain the cross-validated prediction error, which will be used as our performance measure. We repeat this procedure for 30 replicated datasets and summarize the results in \textbf{Figure \ref{fg_HeatAMC_AR_Beta1}} through \textbf{Figure \ref{fg_HeatAMC_GD_Gamma1}}.

\subsection*{\textit{Setting Three: Comparing the HNB and TLNPN Models (with TLNPN Population Data)}}

In this simulation study, we again compare the performance of the TLNPN model to the HNB model, but use the TLNPN model as the true model using Quantitative Microbiome Profiling (QMP) data \citep{vandeputte2017quantitative}. We evaluated the performance of both models under different conditions. We consider both AR or GD correlation structures for the latent correlation matrix. We also used the empirical cdf of both the original (untransformed) data and the square root transformation of the data as the true cdf of the population TLNPN model since the QMP dataset is extreme scale data, and we seek to investigate whether the scale and skewness of the data impacts relative model performance. We also vary the proportion of zeros in the TLNPN data (ZP) to evaluate whether zero-proportion has an effect on relative model performance. Finally, we set the correlation parameter, $\rho$, at different values to evaluate if the dependence of the variables has an impact on relative model performance. For this study, we generated the Gaussian-level variables with a correlation matrix of $\boldsymbol{\Sigma}$, so
$\mathbf{x}_i = (X_{i1},X_{i2},\ldots,X_{i5})^{\top} \sim N_p(\boldsymbol{0}, \boldsymbol{\Sigma})$, $i=1,\dots,n$ where $n=1200$.
Let $\hat F_j$ be the empirical cdf of the $j$th variable of the QMP data. 
We generate data such that $y_{ij}= {\hat F_j}^{-1} \circ \Phi (x_{ij})$, $i=1,\ldots,n$ and $j=1,\ldots,p$ where $p = 5$. In this study, we control for the amount of zeros by subsetting on five variables in the QMP dataset that had the desired zero-proportions, then selecting those to generate the TLNPN data. We then split the population data into five folds, which we used for five-fold cross validation. 
We fitted the TLNPN and HNB models to the training data and generated simulated data from each model, and find the respective Wasserstein distances between the simulated data and the test data. The average between the five-folds is then found and recorded. 

\begin{figure}[hbt!]
\centering
\includegraphics[width=0.5\textwidth]{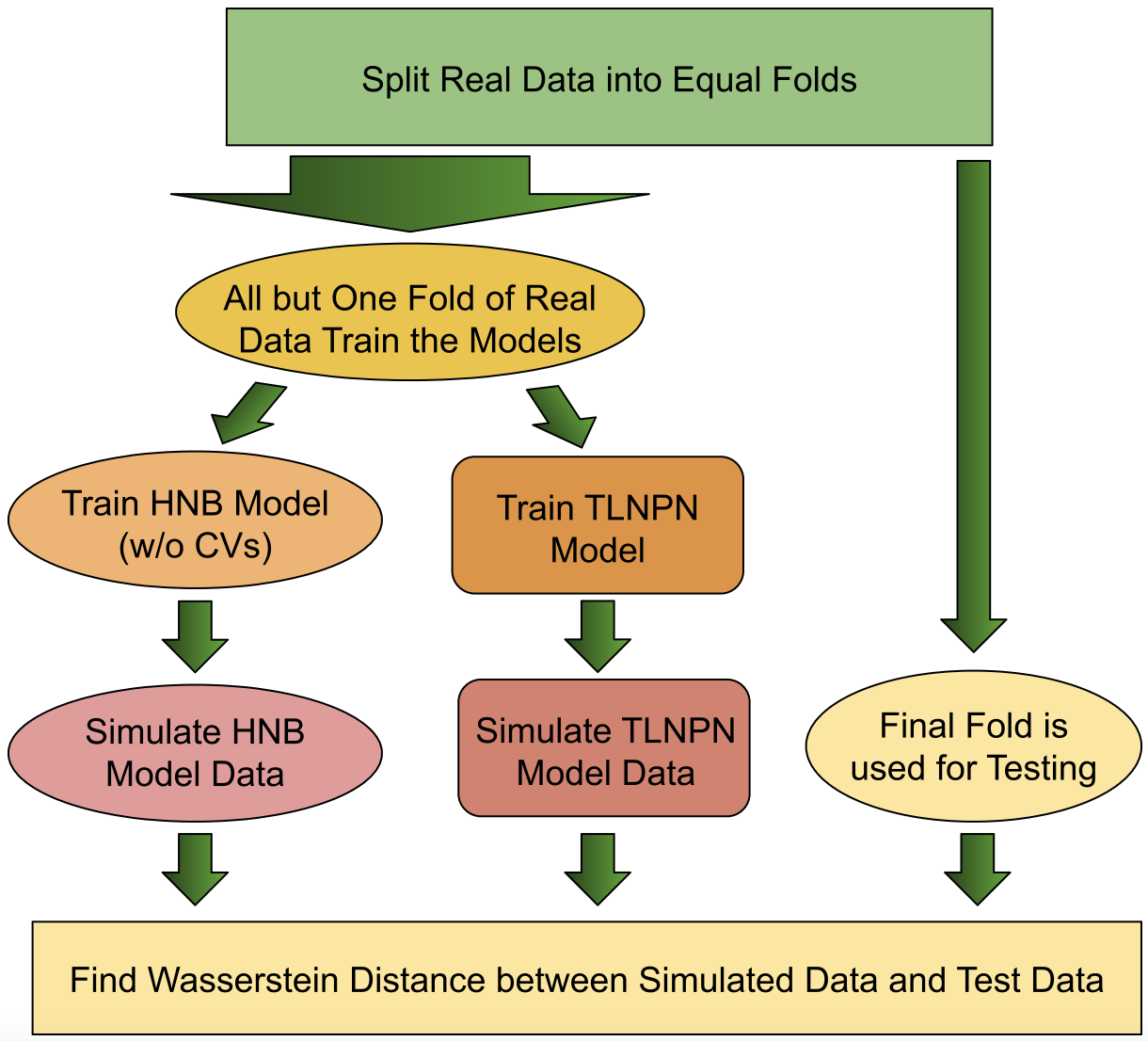}
\caption{Schematic illustration of real data analysis procedure.}
\label{fg_RealDataStudy}
\end{figure}

\subsection*{\textit{Procedure for Real Data Analyses}}

Our real data studies were motivated by datasets from a gut bacteria article \citep{vandeputte2017quantitative} and gene-sequencing data (\textit{https://www.10xgenomics.com}). 
Our analysis process is graphically summarized in \textbf{Figure \ref{fg_RealDataStudy}}. 
We used validation, three-fold for the Quantitative Microbiome Profiling data and five-fold for the gene sequencing data, and we considered 50 random splits. 
We trained the HNB and TLNPN models on all but one fold, then simulated data from those models and found the Wasserstein distance between the simulated data from each model with the final fold. 
We summarize the relative performance of the models using AMC as given by \textbf{Equation \ref{eq_AMC}}.



\section*{RESULTS}

\subsection*{Setting One: Comparing the univariate ZINB and HNB Models}
 
\begin{figure}[!htb]
\centering
\includegraphics[width = 1.0\textwidth]{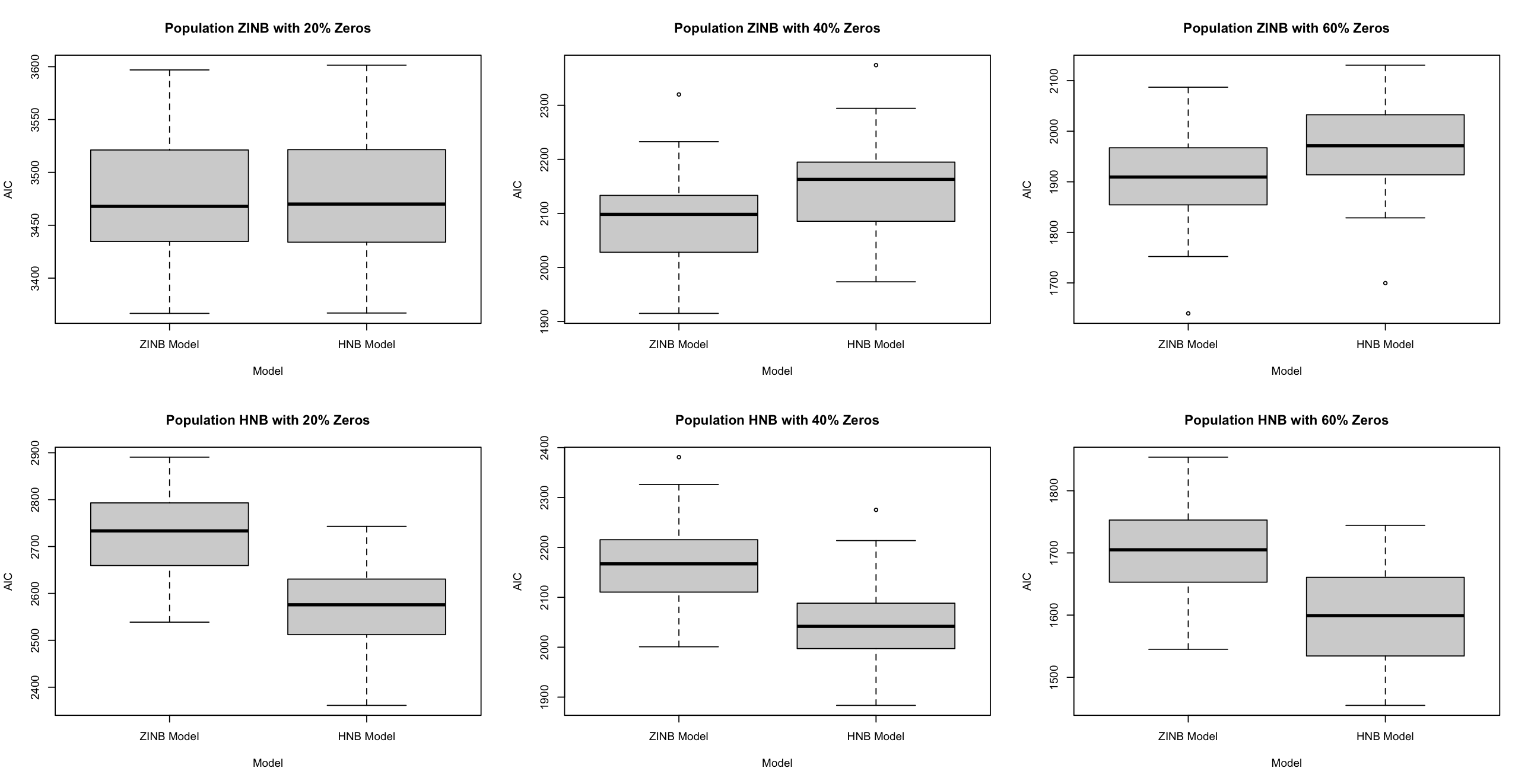}
\caption{The box plots show how the AIC of the ZINB and HNB models compare with one another under varying conditions. On the top row, the population data was generated from the ZINB model, and on the bottom row, the population data was generated from the HNB model. In this case, $\beta_1 = \gamma_1 = 2$. Each column of box plots corresponds to a different proportion of zeros: 20\%, 40\%, and 60\%. We see in the top row that the ZINB model outperforms the HNB model, and on the bottom row, the HNB model outperforms the ZINB model. We see this result because given a low covariate, the ZINB model will predict a sampling zero where the HNB would not.}
\label{fg_ZINBvHNBcv}
\end{figure}

\begin{figure}[!htb]
\centering
\includegraphics[width = 0.5\textwidth]{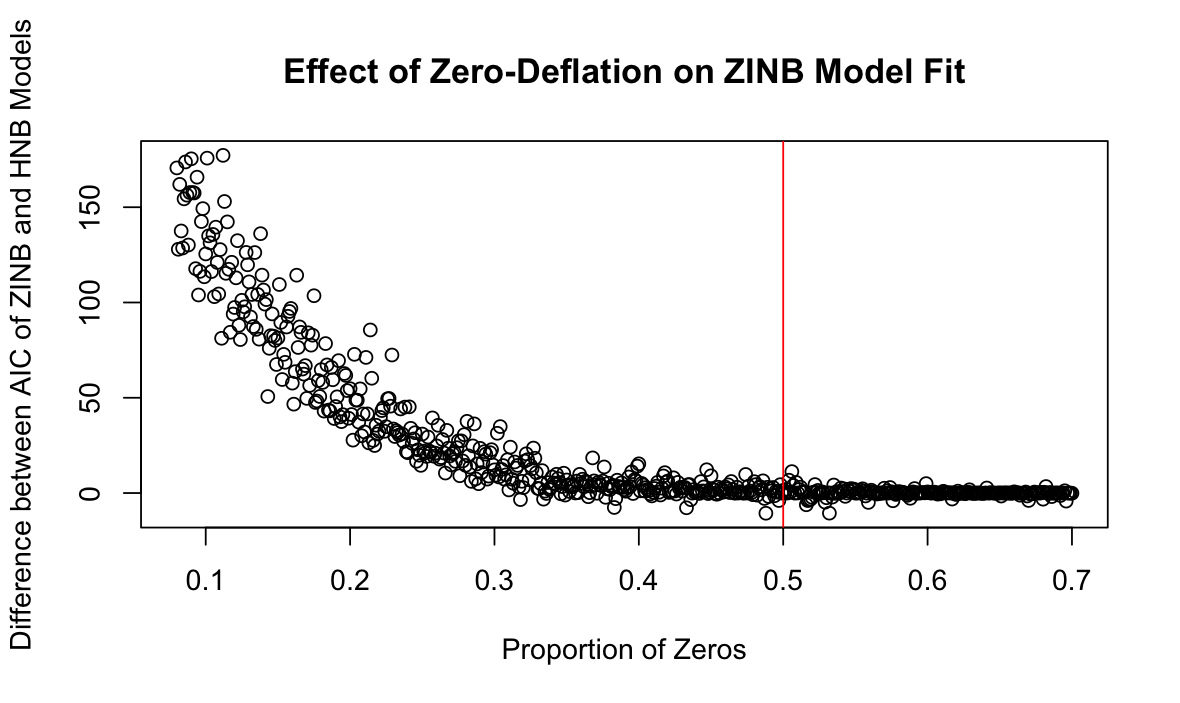}
\caption{In cases of zero-deflation, the HNB far outperformed the ZINB model. We simulated HNB population data under varying levels of zero-deflation and inflation, and fitted both the HNB and ZINB model. We found the difference in AIC between the two models corresponding to each proportion of zeros. Under the standard NB model, the probability that $Y=0$ is 0.5 (the red line).}
\label{fg_ZeroDeflationZINB.png}
\end{figure}

We confirmed previous findings and found that when the proportion of sampling and structural zeros differed significantly and $\beta_1$ and $\gamma_1$ had high values, then the models were sensitive to model misspecification as seen in \textbf{Figure \ref{fg_ZINBvHNBcv}} \citep{feng2021comparison}.
We see that under each proportion of zeros (20\%, 40\%, and 60\%), when $\beta_1=\gamma_1=2$, the true model far outperformed the other in terms of AIC. 
In a second simulation study comparing the HNB and ZINB models, our results show that under conditions of zero deflation—where the probability of $Y=0$ is less than 0.5—the HNB model significantly outperforms the ZINB model in AIC. The difference in AIC values grows exponentially as the proportion of zeros falls below the probability of $Y=0$ under the standard negative binomial distribution, as illustrated in \textbf{Figure \ref{fg_ZeroDeflationZINB.png}}. However, it seems that the ZINB model is able to account for moderate zero-deflation without an impact on model performance.

\subsection*{Setting Two: Comparing the HNB and TLNPN Models (with HNB Population Data)}


\begin{figure}[hbt!]
\centering
    \includegraphics[width=0.49\textwidth]{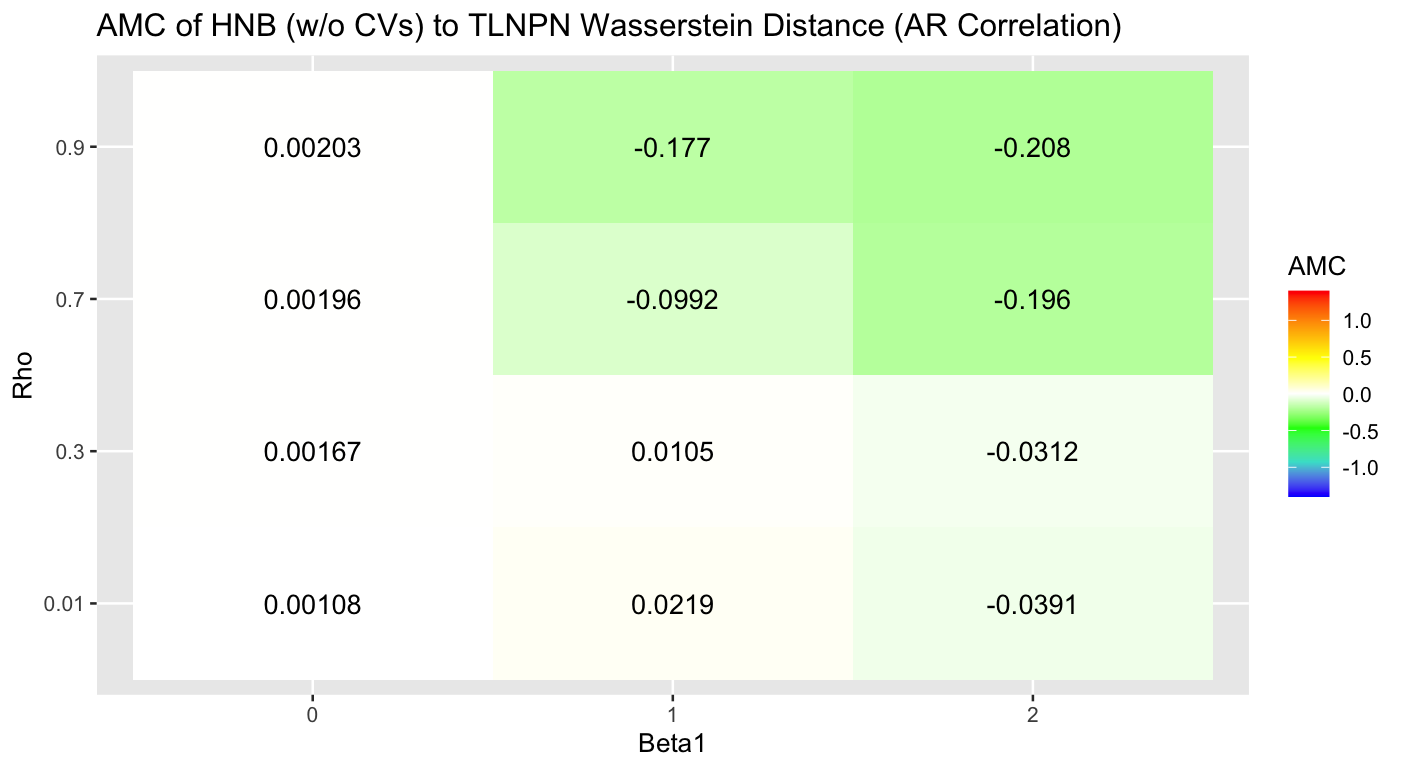}
    \includegraphics[width=0.49\textwidth]{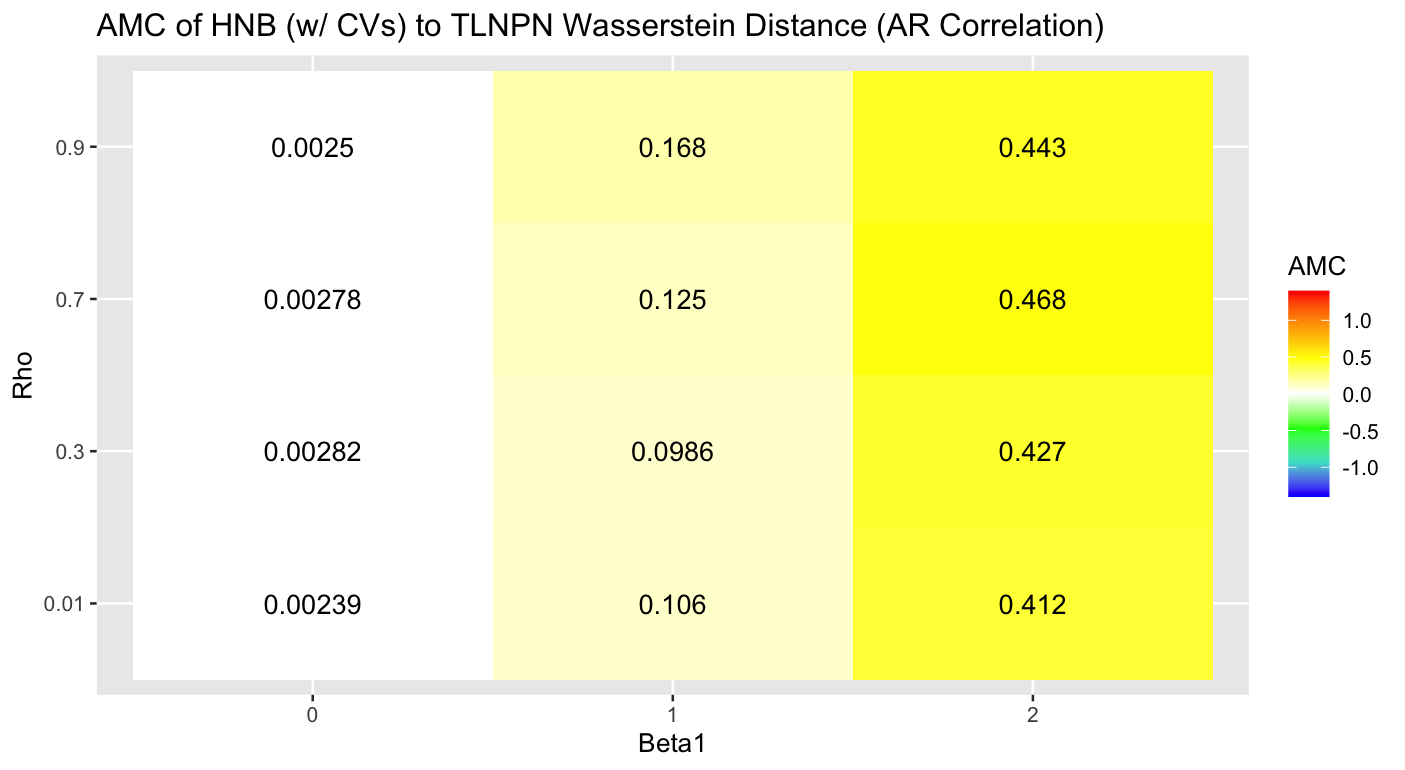}
    \caption{These heatmaps compare the performance of the HNB model to the TLNPN model through our summary measure arithmetic mean change (AMC) given in \textbf{Equation \ref{eq_AMC}}. Positive AMC, represented by warm colors, indicates the HNB model performing better, and negative AMC, represented by cool colors, indicates the TLNPN model performing better. When the HNB model is fitted without covariates, we see that when $\rho$ and $\beta_1$ are high, the TLNPN model outperforms the HNB model (left). However, when the HNB model is fitted with covariates, the HNB model outperforms the TLNPN model most when $\rho$ and $\beta_1$ are high (right).}
    \label{fg_HeatAMC_AR_Beta1}
\end{figure}

We first investigate the interaction between $\beta_1$ and $\rho$ and its impact on relative model performance. 
When we account for whether covariates were included in the fitting of the HNB model, we find starkly different results as shown in \textbf{Figure \ref{fg_HeatAMC_AR_Beta1}}. We see that when covariates aren't considered in the fitting of the HNB model, we see results consistent with our predictions that a high $\rho$ and high $\beta_1$ parameter would improve the relative performance of the TLNPN model. We also see that when $\beta_1=0$ or $\rho = 0.01$, the models perform nearly the same. However, when covariates are considered in fitting the HNB model, we see notably different results. Still, the TLNPN and HNB models perform nearly the same when $\beta_1 = 0$ with the AMC of the Wasserstein distances being approximately 0 regardless of $\rho$. However, as $\beta_1$ increases, the relative performance of the TLNPN model to the HNB model worsens. Furthermore, it seems that the as $\rho$ increases, the TLNPN model performs worse against the HNB model fitted with covariates. 
We conducted a follow-up simulation study to investigate this trend. 
\begin{figure}[htb!]
	\centering
    \includegraphics[width=0.5\textwidth]{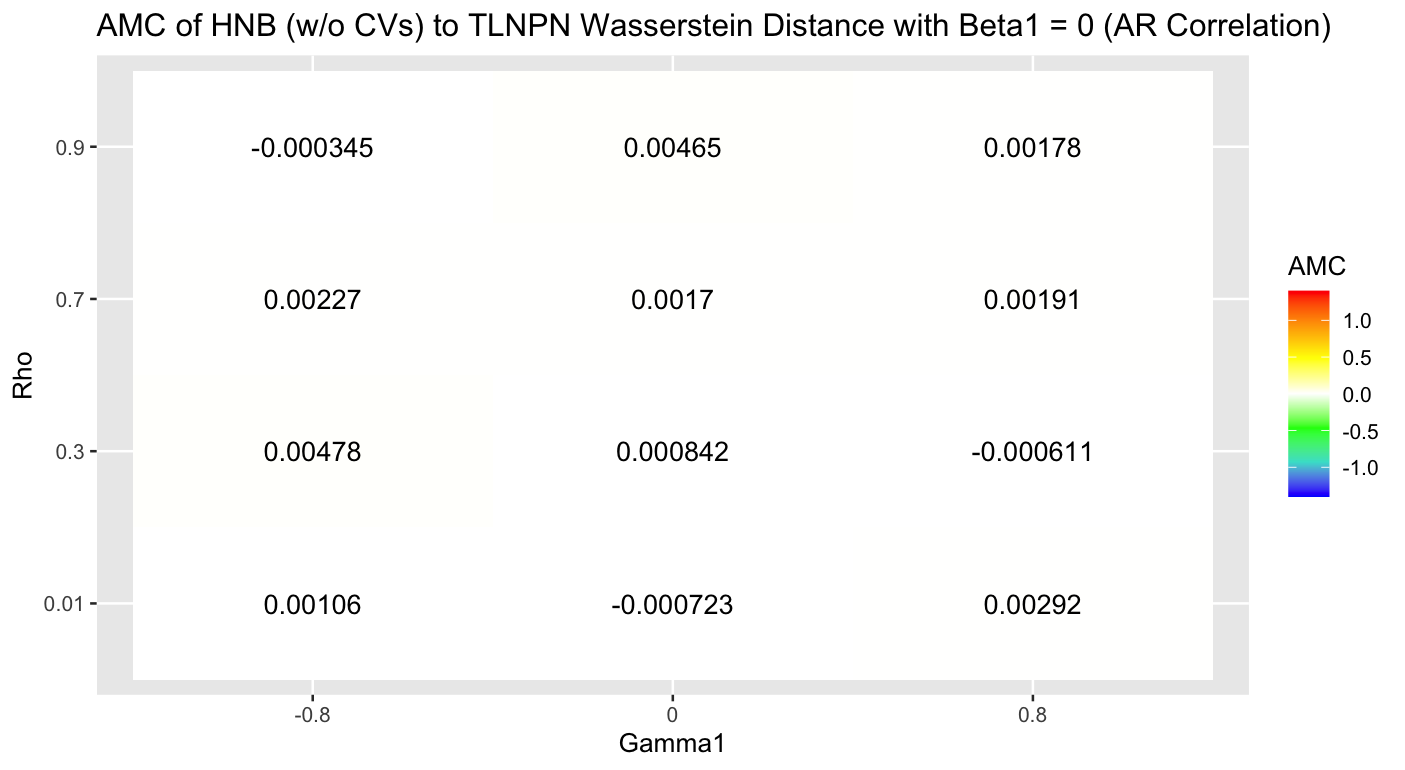}
    \caption{When $\beta_1=0$, the HNB model and TLNPN model perform the same. We see this result because when $\beta_1=0$, then the HNB variables are practically independent (the $\gamma_1$ parameter could still incur a minor amount of dependence between the variables). Therefore, the TLNPN model doesn't have an advantage in fitting the multivariate distribution. }
     \label{fg_HeatAMC_AR_Gamma1_B0_ncv}
\end{figure}

\begin{figure}[htb!]
	\centering
    \includegraphics[width=0.49\textwidth]{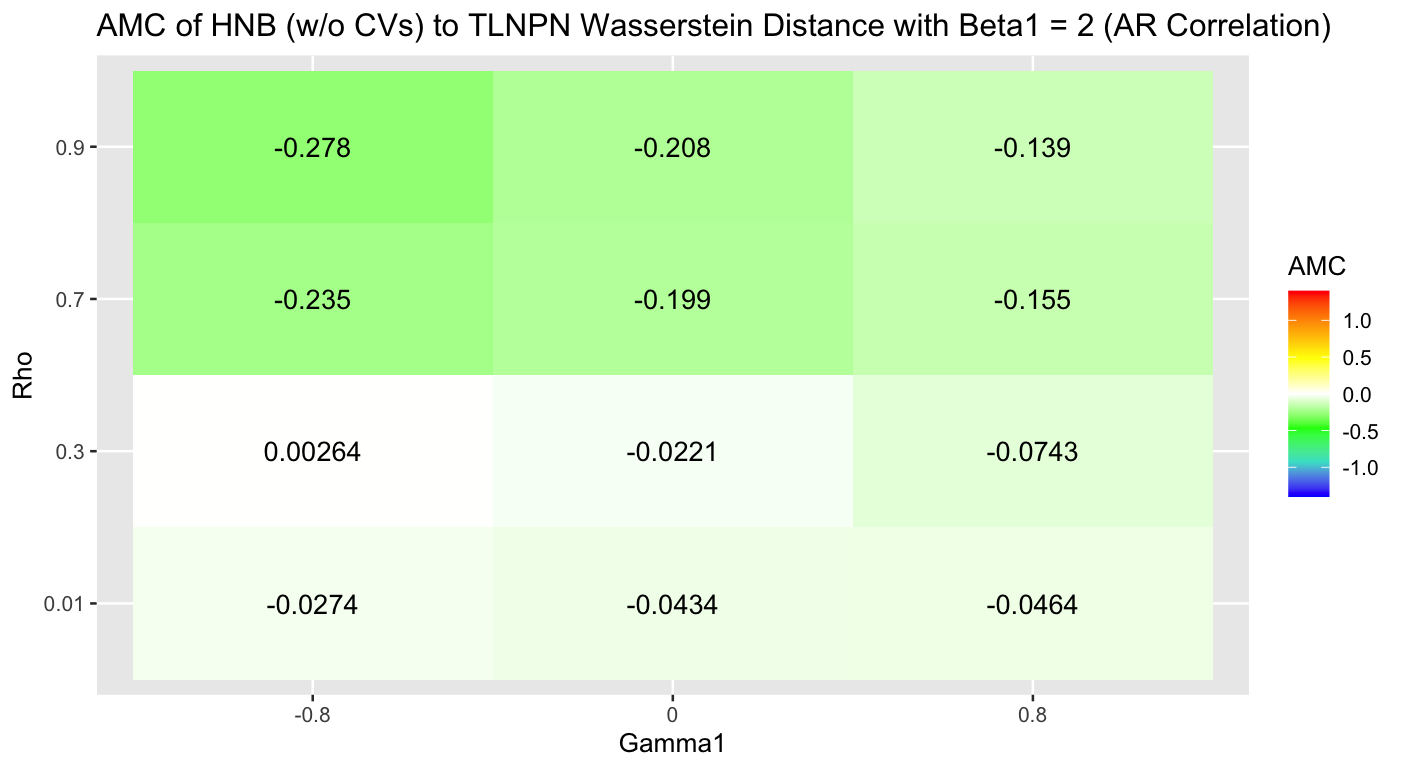}
    \includegraphics[width=0.49\textwidth]{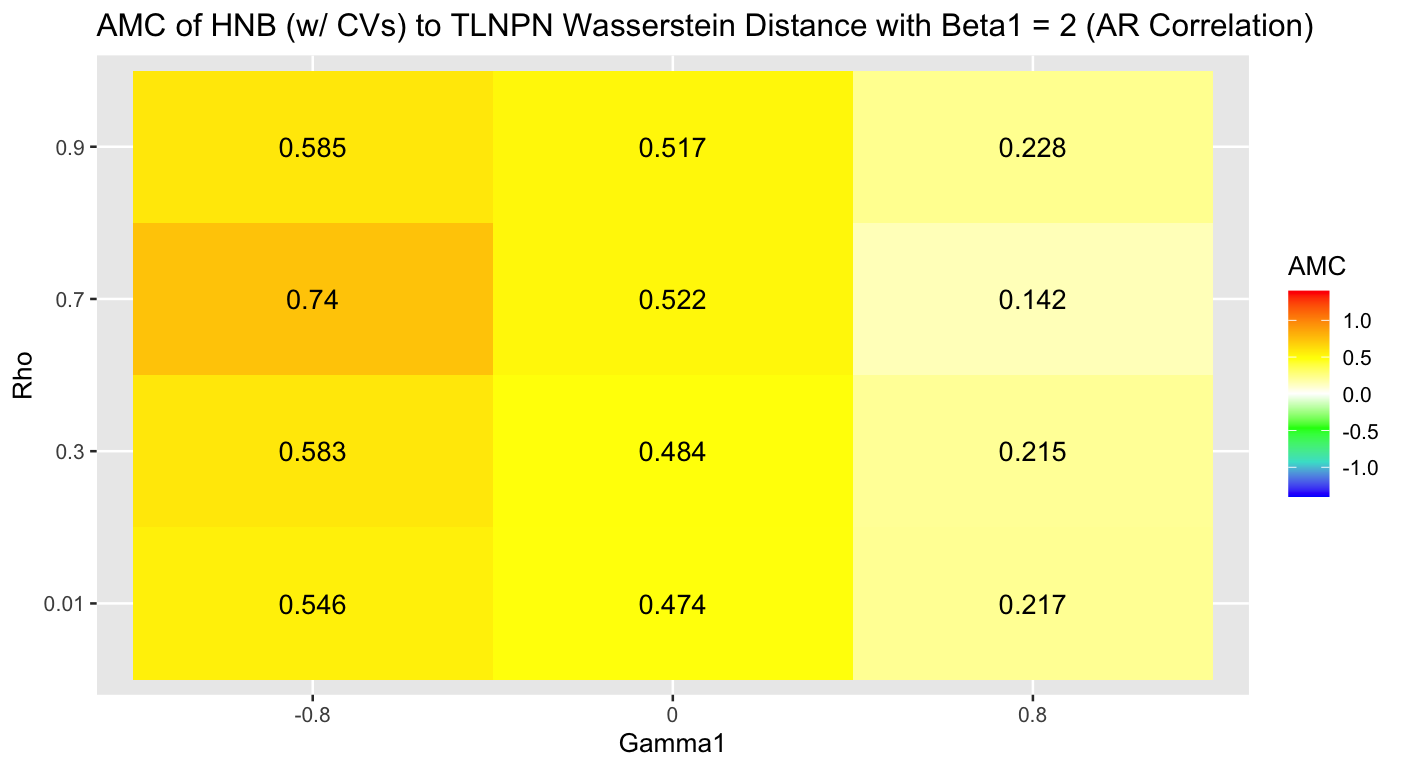}
	\caption{The heatmaps show the relative performance of the TLNPN model to the HNB model under different values of $\rho$ and $\gamma_1$. We see that when covariates aren't considered when fitting the HNB model, then the TLNPN model performs best relative to the HNB model when $\gamma_1=-0.8$ and $\rho=0.9$ (left). This is a result of the stronger dependence among the variables that results when $\gamma_1=-0.8$. We also see that when covariates are considered in fitting the HNB model, then the TLNPN model performs worse relative to the HNB model as $\rho$ increases and $\gamma_1$ decreases (right).}
     \label{fg_HeatAMC_AR_Gamma1_B2_ncv}
\end{figure}

\begin{figure}[hbt!]
	\centering
    \includegraphics[width=0.85\textwidth]{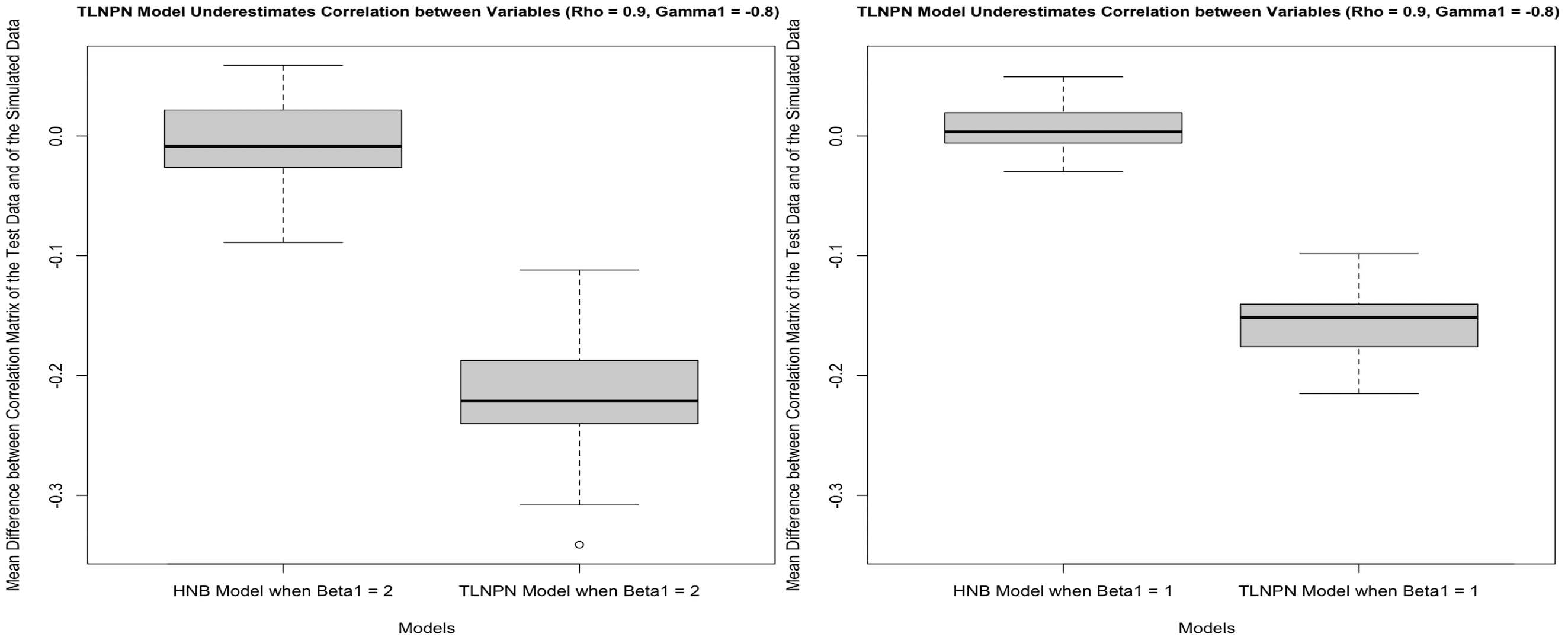}
	\caption{These box plots show that as $\beta_1$ increases, the TLNPN increasingly underestimates correlation between variables. In this figure, the HNB model was fitted with covariates. We suspect this occurs because the latent Gaussian variables in the TLNPN model must have a correlation matrix such that the joint distribution will have data points with variables at high values and variables equal to 0 on account of the zero-inflation.}
     \label{fg_BoxCord}
\end{figure}

 The follow-up simulation study followed the same process as the first except we also measured the marginal Wasserstein distance between the test data and the simulated data and the correlation matrix of the test data and the correlation matrix simulated data. In this study, we considered the HNB model fitted with covariates. We found that as $\beta_1$ increases from $1$ to $2$ (when $\rho = 0.9$), the TLNPN model will further underestimate the correlation between the zero-inflated variables as shown in \textbf{Figure \ref{fg_BoxCord}}.
 \begin{figure}[hbt!]
	\centering
    \includegraphics[width = 0.98\textwidth]{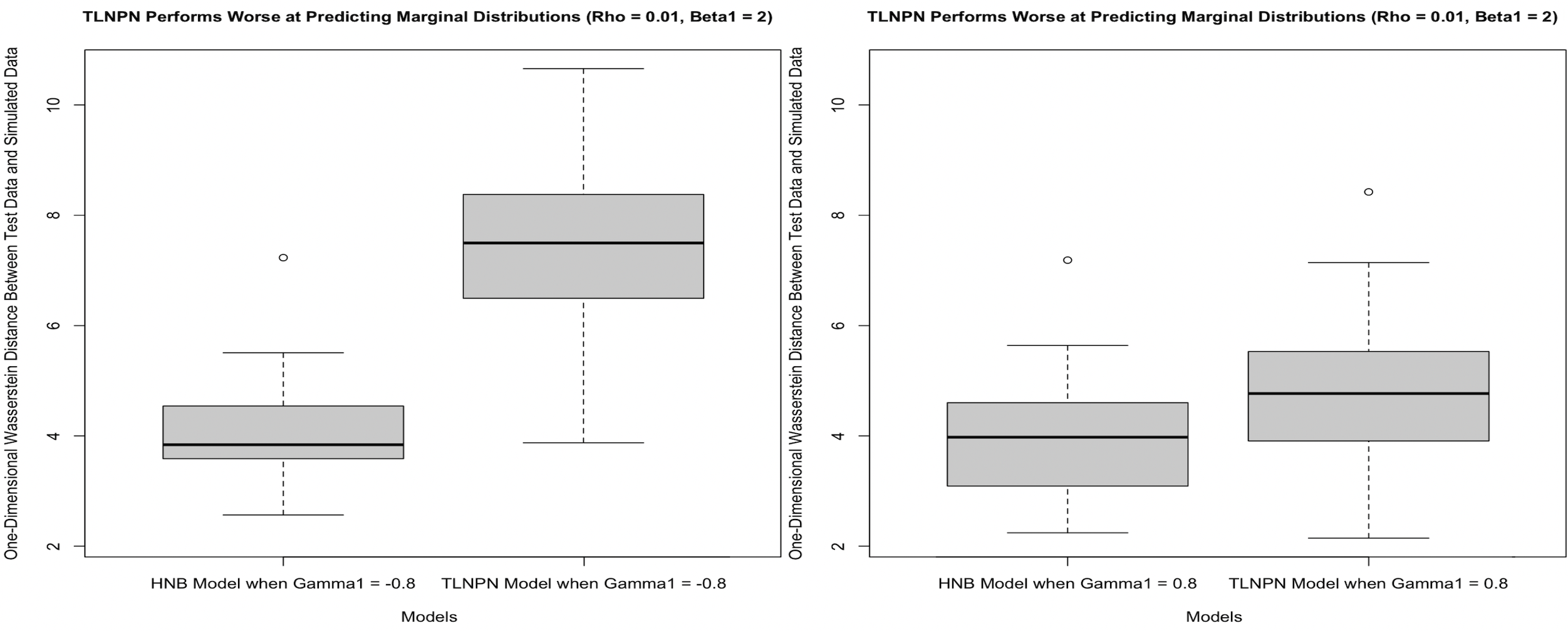}
	\caption{These box plots show the impact that $\gamma_1$ has on the marginal performance of the TLNPN model. We see that as $\gamma_1$ increases, the Wasserstein distance between the TLNPN data and test data decreases. This trend is a result of higher, more extreme values become less probable as $\gamma_1$ increases, which the TLNPN model struggles to predict. The HNB model was fitted with covariates, and we see that the marginal Wasserstein distance stays approximately the same as $\gamma_1$ increases. }
     \label{fg_Box1DwdistHNBvTLNPNB2}
\end{figure}
  We also found that the one-dimensional Wasserstein distances of the marginal TLNPN data to the marginal test data were greater than that of those generated by HNB data as displayed in \textbf{Figure \ref{fg_Box1DwdistHNBvTLNPNB2}}. 

\begin{figure}
\centering
\includegraphics[width=0.6\textwidth]{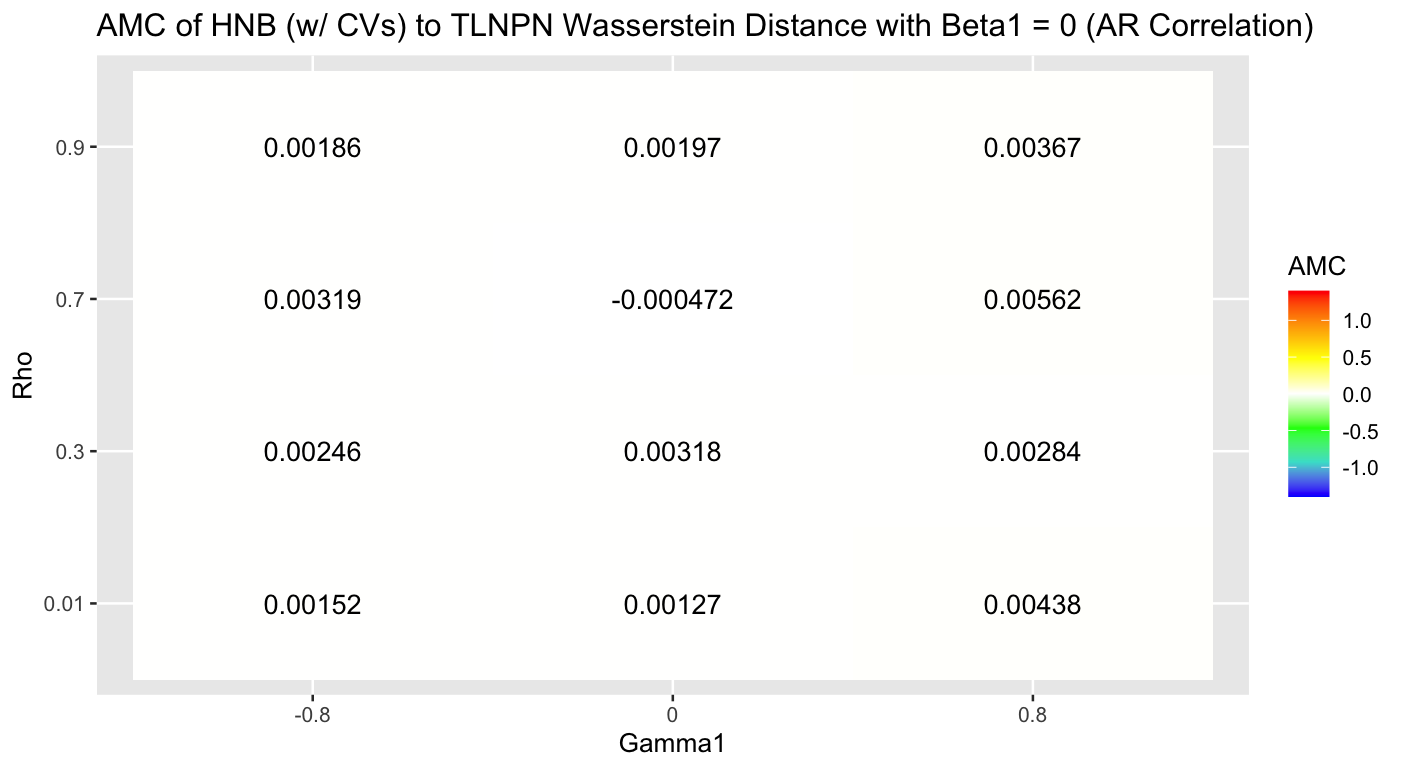}
\caption{This shows no difference in performance between the HNB and TLNPN models when $\beta_1 = 0$}
\label{figures/HeatAMC_AR_Gamma1_B0_cv}
\end{figure}

We found that the zero-proportion of the variables, controlled by $\gamma_0$, did not have an impact on relative model performance between the HNB and TLNPN models.
Furthermore, we also investigated the interaction between $\rho$ and $\gamma_1$ with regards to the relative model fit between the TLNPN and HNB models when controlling for $\beta_1$. 
We see in \textbf{Figure \ref{fg_HeatAMC_AR_Gamma1_B0_ncv}} and \textbf{Figure \ref{figures/HeatAMC_AR_Gamma1_B0_cv}} that regardless of whether covariates are included in the HNB model fitting, when $\beta_1 = 0$, $\gamma_1$ has no impact on the relative performance of the TLNPN model. 
When $\beta_1 = 2$, we see a trend in \textbf{Figure \ref{fg_HeatAMC_AR_Gamma1_B2_ncv}}. When covariates are not considered, the TLNPN model performs best when $\gamma_1 = -0.8$ and when $\rho = 0.9$. Additionally, as $\rho$ increases, the performance of the TLNPN model against the HNB model improves. 
When covariates are considered, the trend is reversed where the TLNPN performs worse when $\gamma_1 = -0.8$ and $\rho$ is high. We conducted a follow-up simulation study to investigate this trend.

\begin{figure}[hbt!]
\centering
\includegraphics[width = 0.9\textwidth]{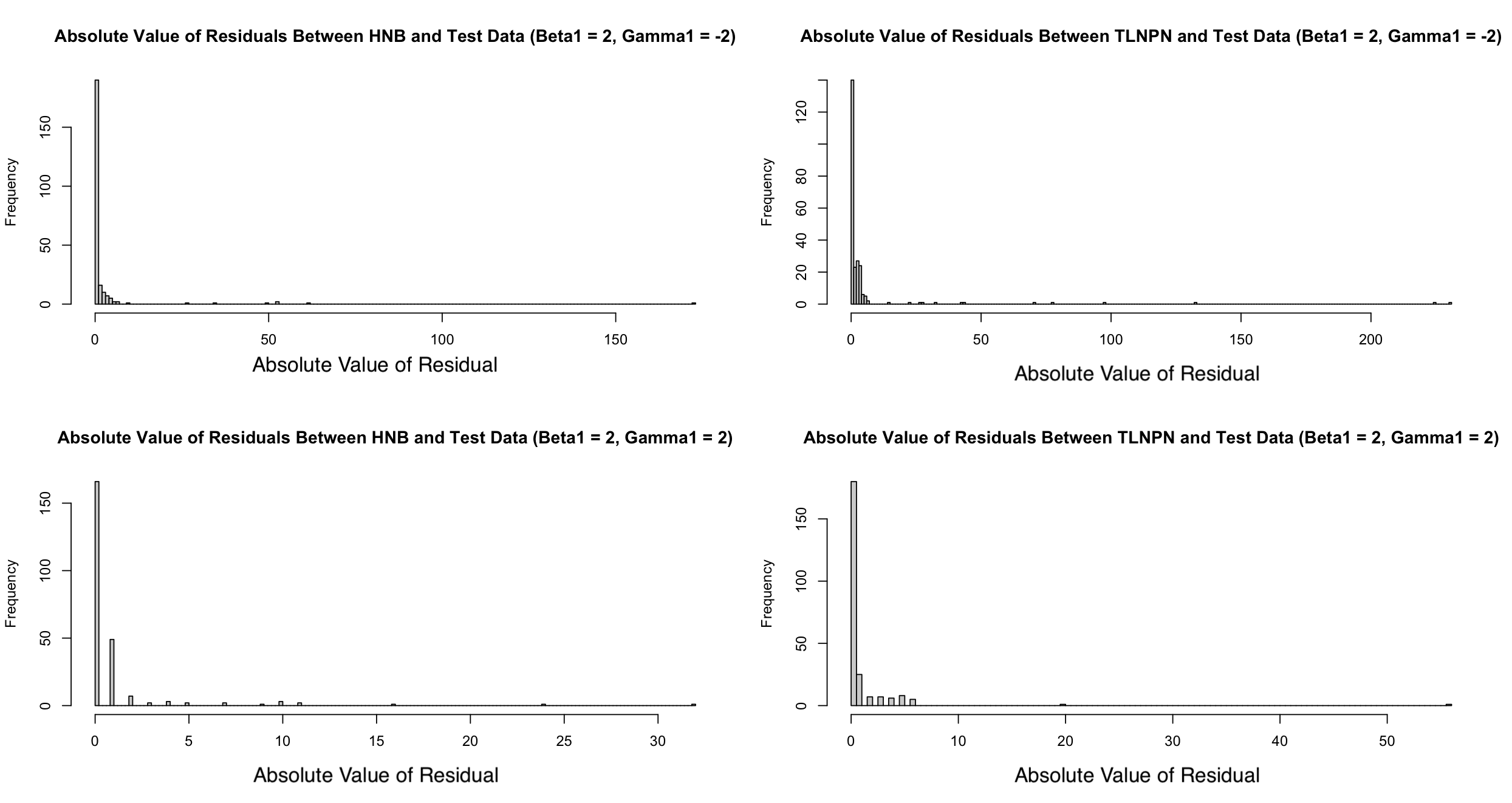}
\caption{These histograms display how $\gamma_1$ affects the marginal performance of the HNB and TLNPN models in the first and second column respectively. The first row shows the difference between the sorted values of the simulated and test data when $\gamma_1=-2$, and the second row shows the same for $\gamma_1=2$. As $\gamma_1$ increases, the graphs show the residuals decreasing, particularly for the TLNPN data, which occurs because when $\gamma_1$ increases, given $\beta_1 > 0$, larger values become less likely since as the covariate increases, both $\mu_{ij}$ and $\pi_{H,ij}$ increase.}
\label{fg_HistResidualTLNPNvHNB}
\end{figure}

From our previous follow-up simulation study, we see that as $\gamma_1$ increases, the one-dimensional Wasserstein distance between the TLNPN data and the test data decreases as seen in \textbf{Figure \ref{fg_Box1DwdistHNBvTLNPNB2}}. 
To explain the differences in the one-dimensional Wasserstein distances, we conducted another follow-up simulation study in which we compared the distributions of the test data to the HNB and TLNPLN simulated data when $\gamma_1=-2$ and when $\gamma_1=2$, and the results are displayed in \textbf{Figure \ref{fg_HistResidualTLNPNvHNB}}. We see that when $\gamma_1 = -2$, we find much higher residuals between the TLNPN and test data; however, when $\gamma_1 = 2$, these residuals decrease dramatically. 

The results of this simulation study presented thus far have been generated from covariates following an AR correlation structure; we found similar results from covariates generated from the GD correlation structure. Again, note that in a GD correlation structure, as $\rho$ decreases, the correlation (in an absolute sense) between the covariates tends to increase. 

\begin{figure}[hbt!]
\centering
\includegraphics[width = 0.49\textwidth]{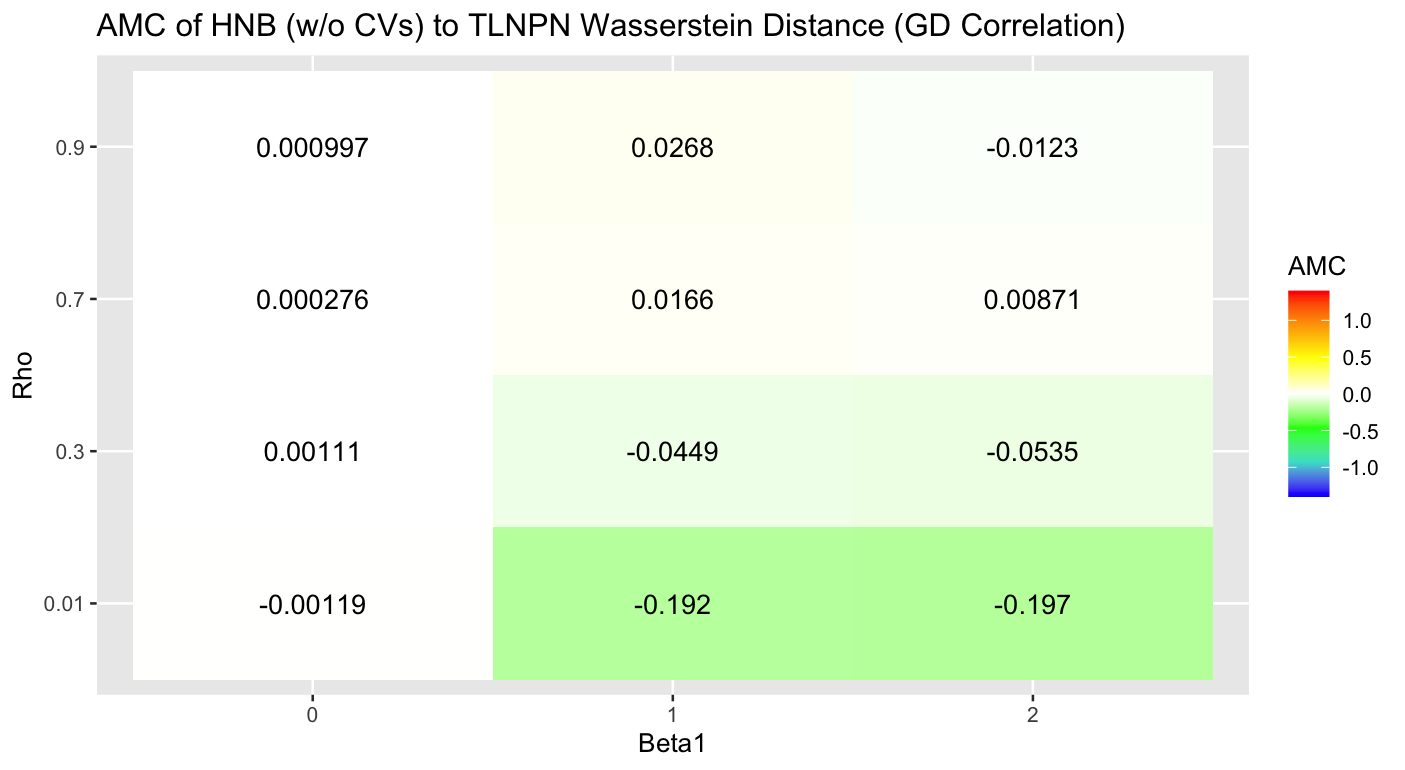}\includegraphics[width = 0.49\textwidth]{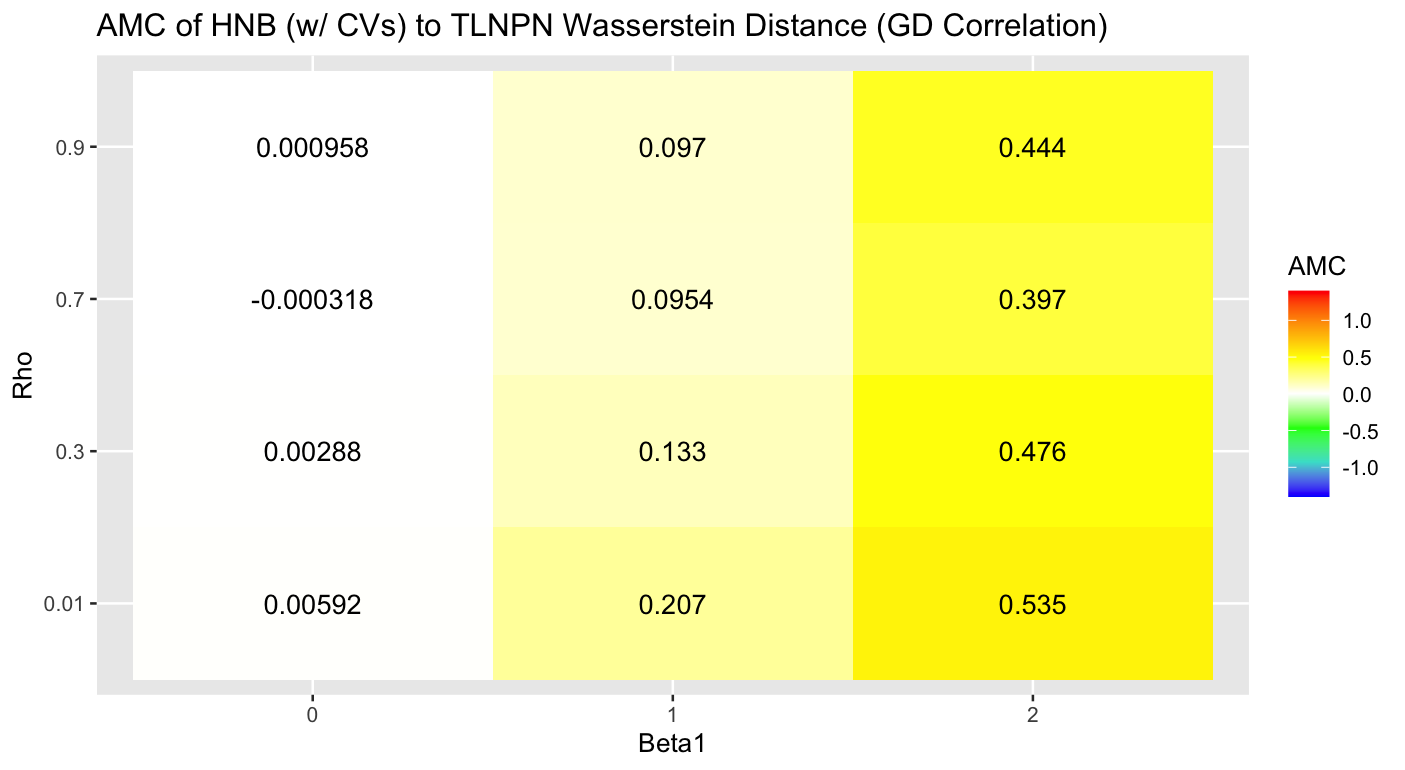}
\caption{We see in the first heat map that the TLNPN model outperforms the HNB model (fitted without covariates) most when $\rho=0.01$ and $\beta_1=2$. This occurs because under the GD correlation structure, when $\beta_1$ is large and $\rho$ is small, there is a stronger dependence among the zero-inflated variables. In the second heat map, when the HNB model is fitted with covariates, we see the opposite trend for the same reasons as under the AR correlation structure. }
\label{fg_HeatAMC_GD_Beta1_ncv}
\end{figure}

We see in \textbf{Figure \ref{fg_HeatAMC_GD_Beta1_ncv}}, that again, there is an interaction between $\rho$ and $\beta_1$ such that when $\rho$ is small and $\beta_1$ is large, the TLNPN model outperforms the HNB model fitted without covariates. We again see similar results when the HNB model is fitted with covariates where the HNB model outperforms the TLNPN model most when $\beta_1=2$ and $\rho=0.01$. For both scenarios, we see that when $\beta_1 = 0$, the models perform nearly identically. 

\begin{figure}[hbt!]
\centering
\includegraphics[width = 0.49\textwidth]{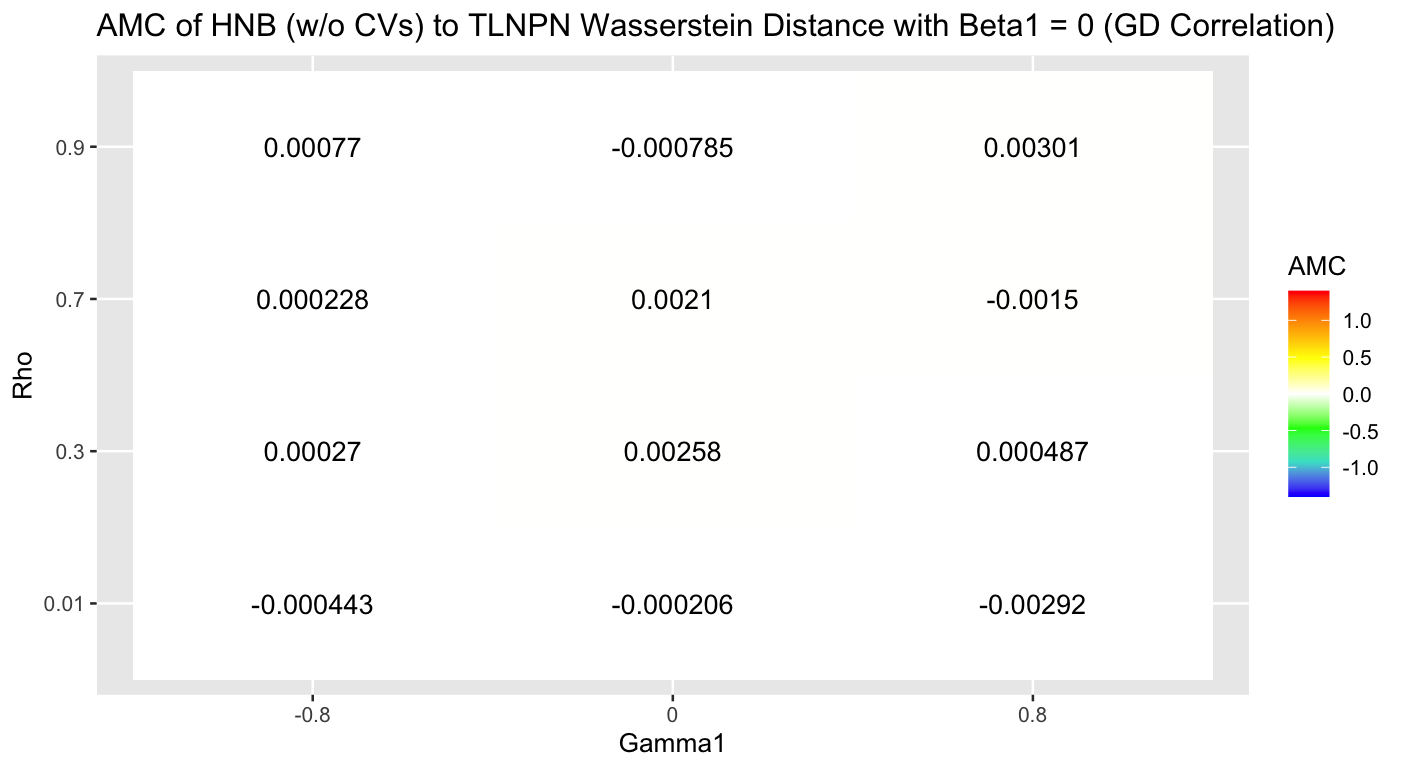}\includegraphics[width = 0.49\textwidth]{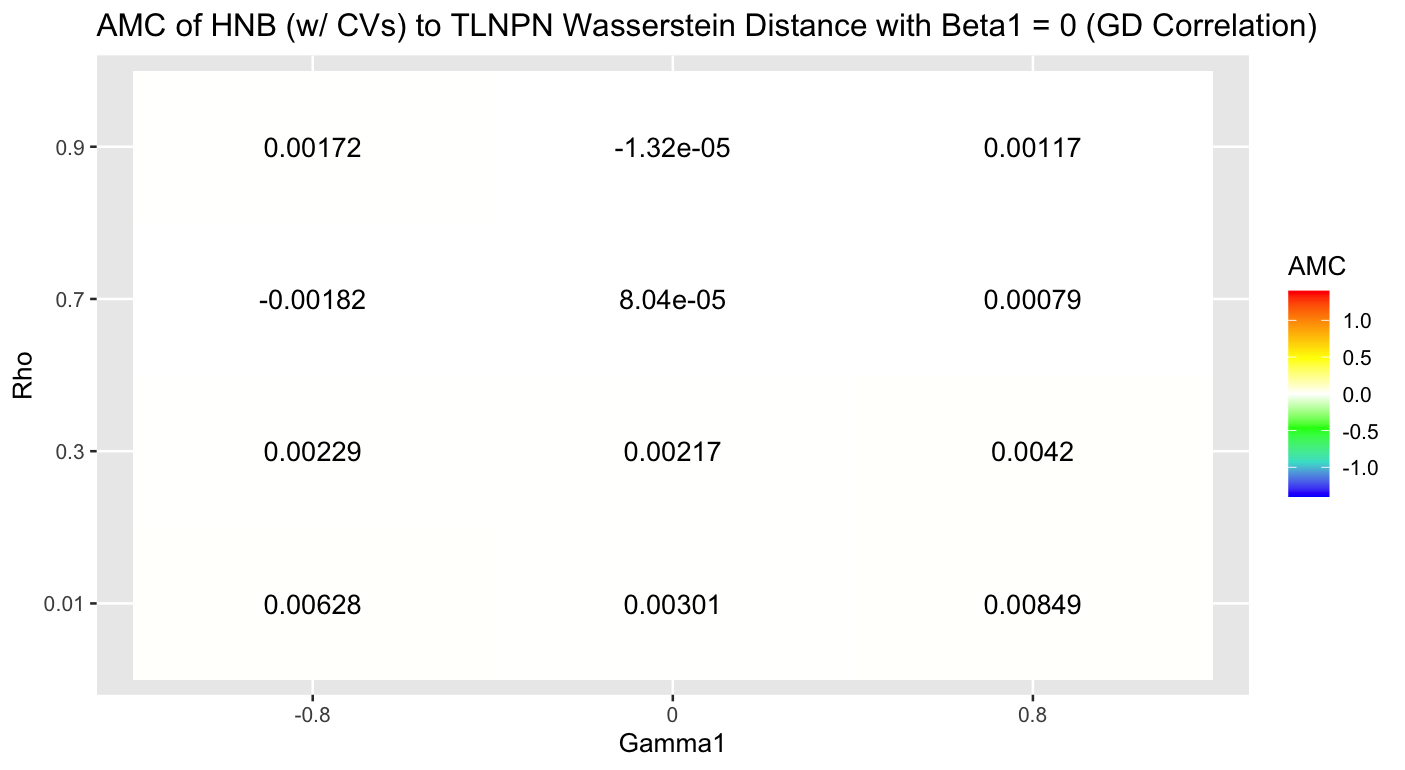}
\includegraphics[width = 0.49\textwidth]{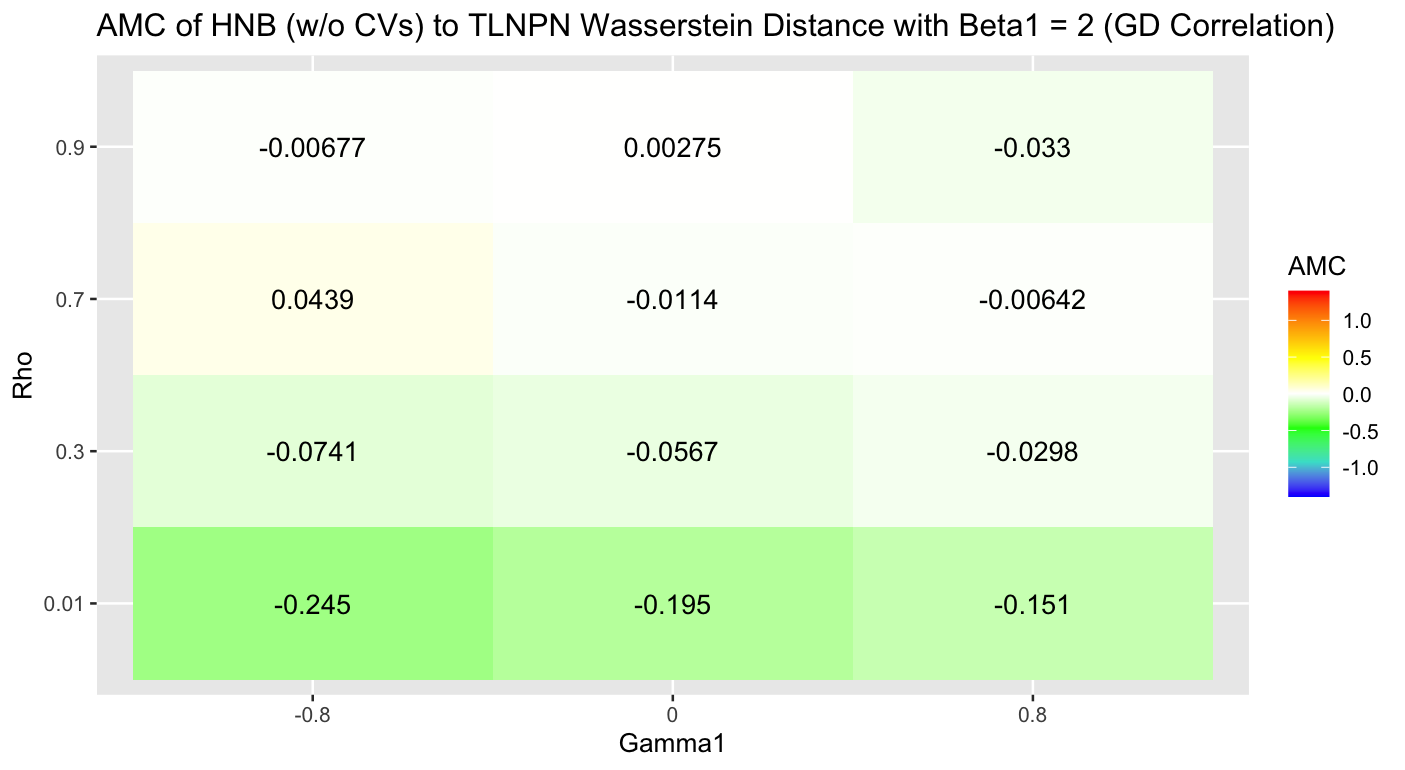}\includegraphics[width = 0.49\textwidth]{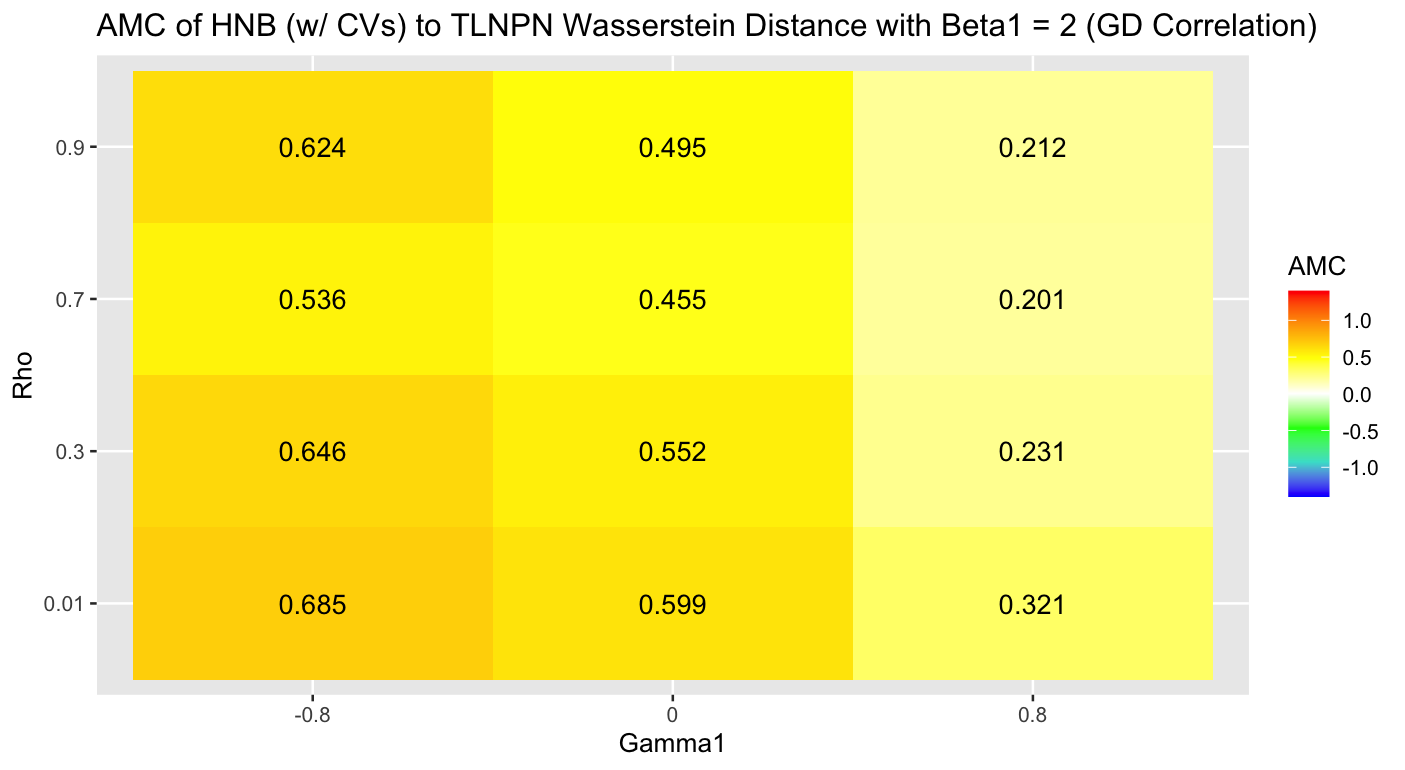}
\caption{On the left column of heat maps, we compare the TLNPN model to the HNB model fitted without covariates. On the right column, we compare the TLNPN model to the HNB model fitted with covariates. On the top row, we consider when $\beta_1=0$, and on the bottom row, we consider when $\beta_1=2$. We see on the top row, that the TLNPN model and HNB models perform nearly identically when $\beta_1=0$ as the variables are practically independent and have much less variance compared to when $\beta_1=2$. We see on the bottom left heat map that the TLNPN model outperforms the HNB model fitted without covariates most when $\rho = 0.01$ and $\gamma_1=-0.8$ as a result of the increased dependence among the variables. We see in the bottom right heat map that when the HNB model is fitted with covariates, the opposite is true. }
\label{fg_HeatAMC_GD_Gamma1}
\end{figure}
As with the AR correlation structure, we again see in \textbf{Figure \ref{fg_HeatAMC_GD_Gamma1}} an interaction between $\beta_1$ and $\gamma_1$ where the effect of $\gamma_1$ only becomes clear when $\beta_1 \ne 0$ since the models perform nearly identically when $\beta_1=0$ regardless of $\rho$ or $\gamma_1$. When $\beta_1=2$, we begin to see a familiar pattern. When covariates are not considered when fitting the HNB model, the TLNPN model outperforms the HNB model most when $\rho=0.01$ and $\gamma_1=-0.8$ where as $\rho$ and $\gamma_1$ decrease, the better the TLNPN model performs relative to the HNB model. However, when covariates are considered when fitting the HNB model, the relative performance of the TLNPN model declines as $\rho$ and $\gamma_1$ decrease.

\subsection*{Setting Three: Comparing the HNB and TLNPN Models (with TLNPN Population Data)}

\begin{figure}[t!]
    \centering
    \includegraphics[width=0.49\textwidth]{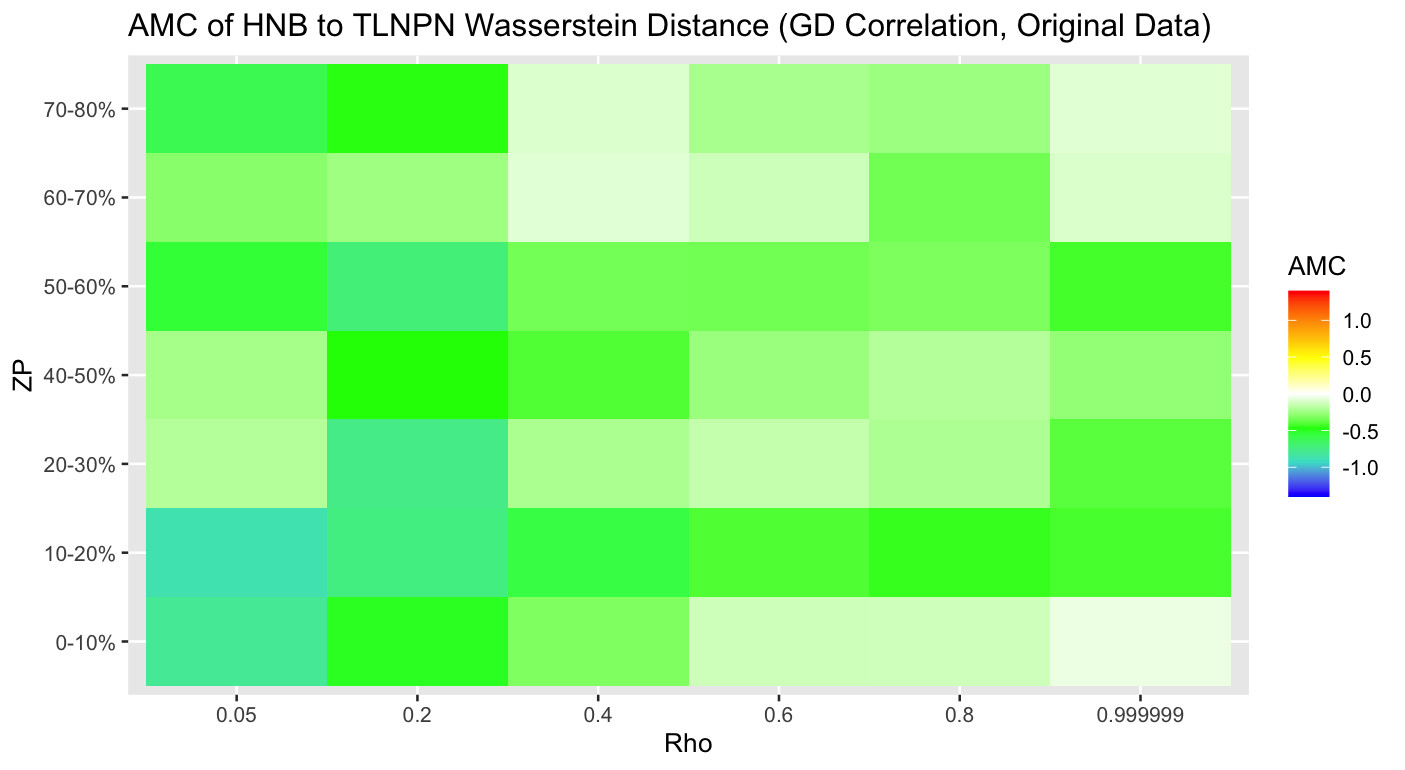}
    \includegraphics[width=0.49\textwidth]{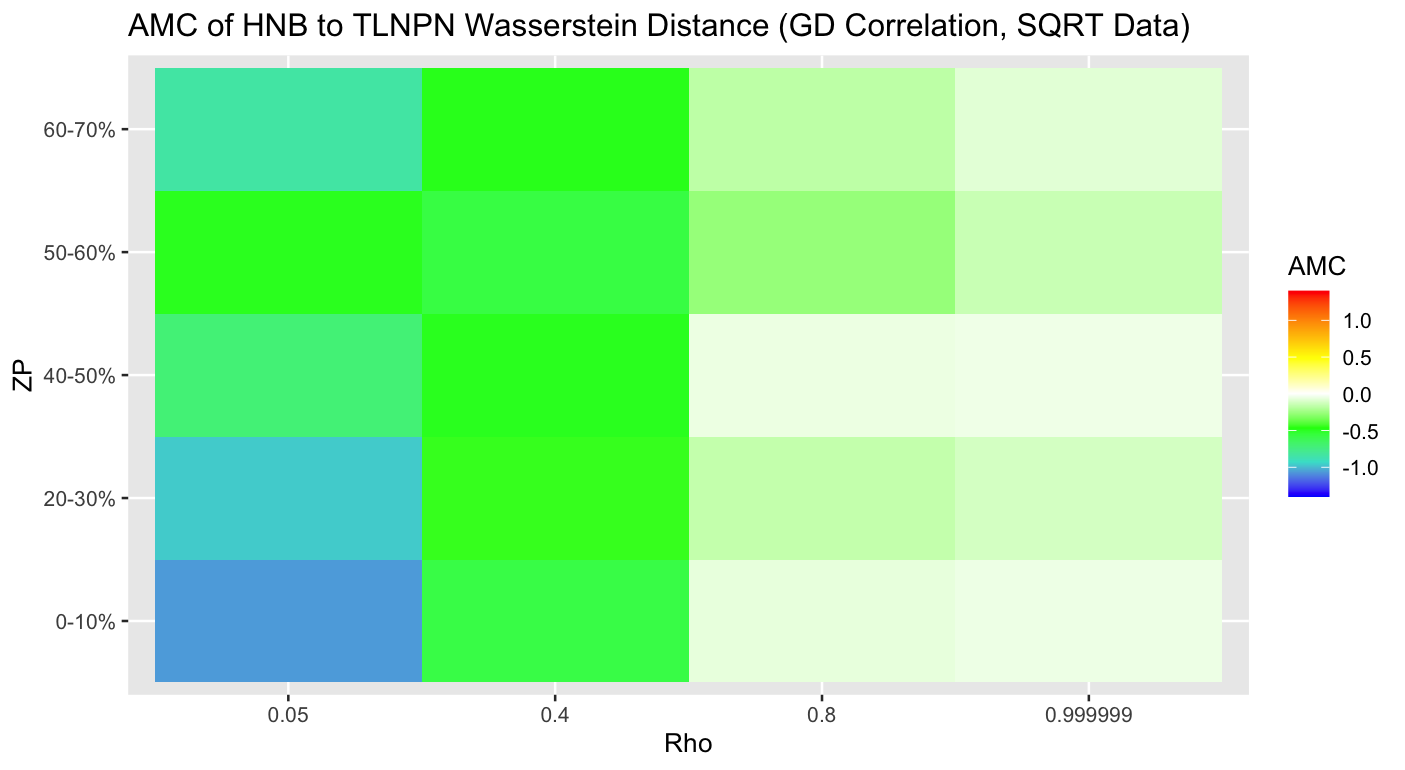}
    \caption{In these heat maps, we compare the performance of the TLNPN and HNB models under TLNPN population data. On the left, the TLNPN model was trained on the untransformed QMP data. On the right, the TLNPN population model was trained on the square root of the QMP data. We see that in both cases, the TLNPN universally performs better regardless of the zero-proportion of the data or $\rho$. Although there's no clear pattern in the first heat map, we see in the second heat map, a clear pattern emerge: as $\rho$ decreases, the TLNPN model improves its performance relative to the HNB model. }
    \label{fg_HeatAMC_ZP_GDod}
\end{figure}


In our third simulation study, we investigate the impact that the zero-proportion and correlation of the TLNPN variables had on relative model performance. 
Our results for the GD correlation structure are presented in \textbf{Figure \ref{fg_HeatAMC_ZP_GDod}}. We see that when evaluating the models under the TLNPN variables distributed as the original, untransformed data, there is no clear pattern; the TLNPN model outperforms the HNB model in all cases, but it's not clear how zero-proportion or $\rho$ impacts the AMC of the HNB Wasserstein distance to the TLNPN Wasserstein distance. However, when we use the square root of the data, we find a much clearer pattern. The lower $\rho$ is when using the GD correlation structure, the better the TLNPN model performs relative to the HNB model.  

\begin{figure}[hbt!]
\centering
\includegraphics[width=0.49\textwidth]{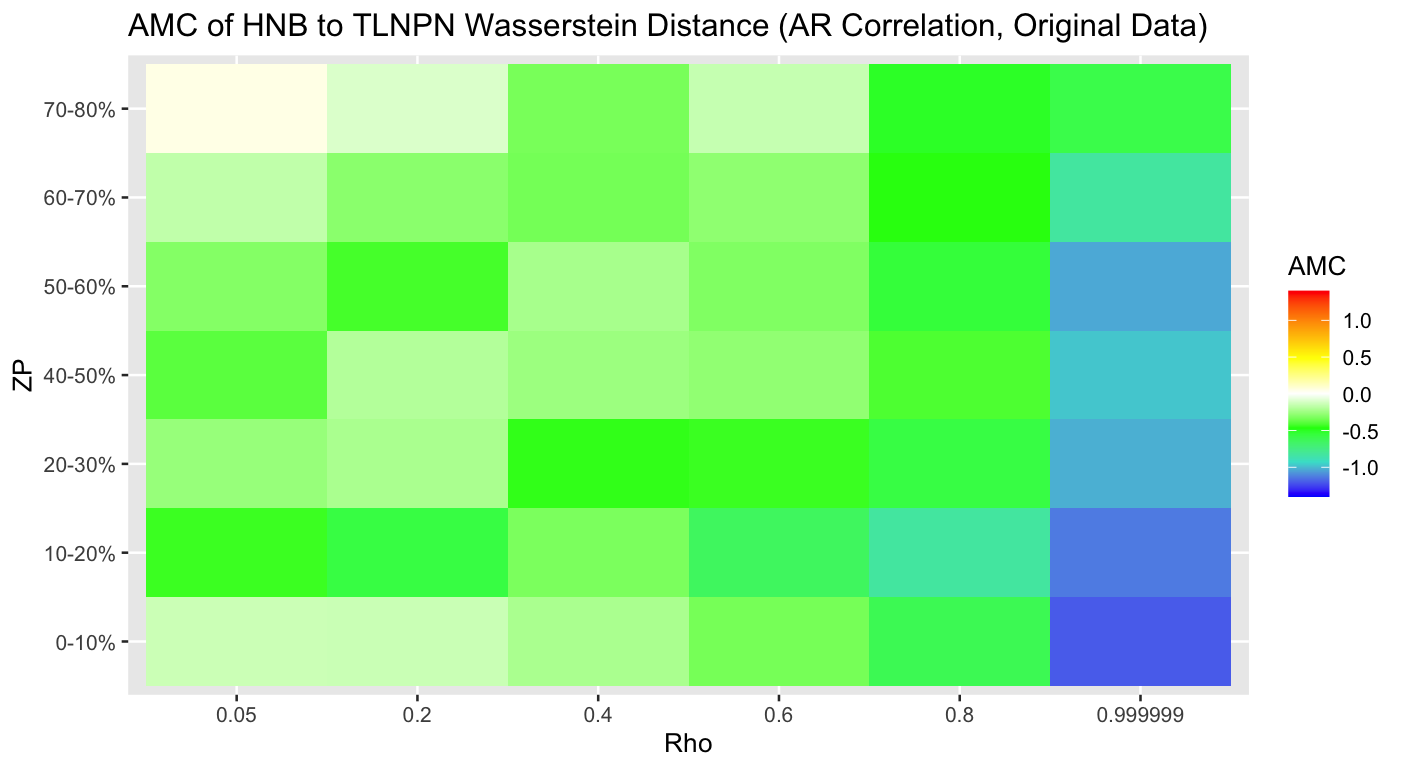}
\includegraphics[width=0.49\textwidth]{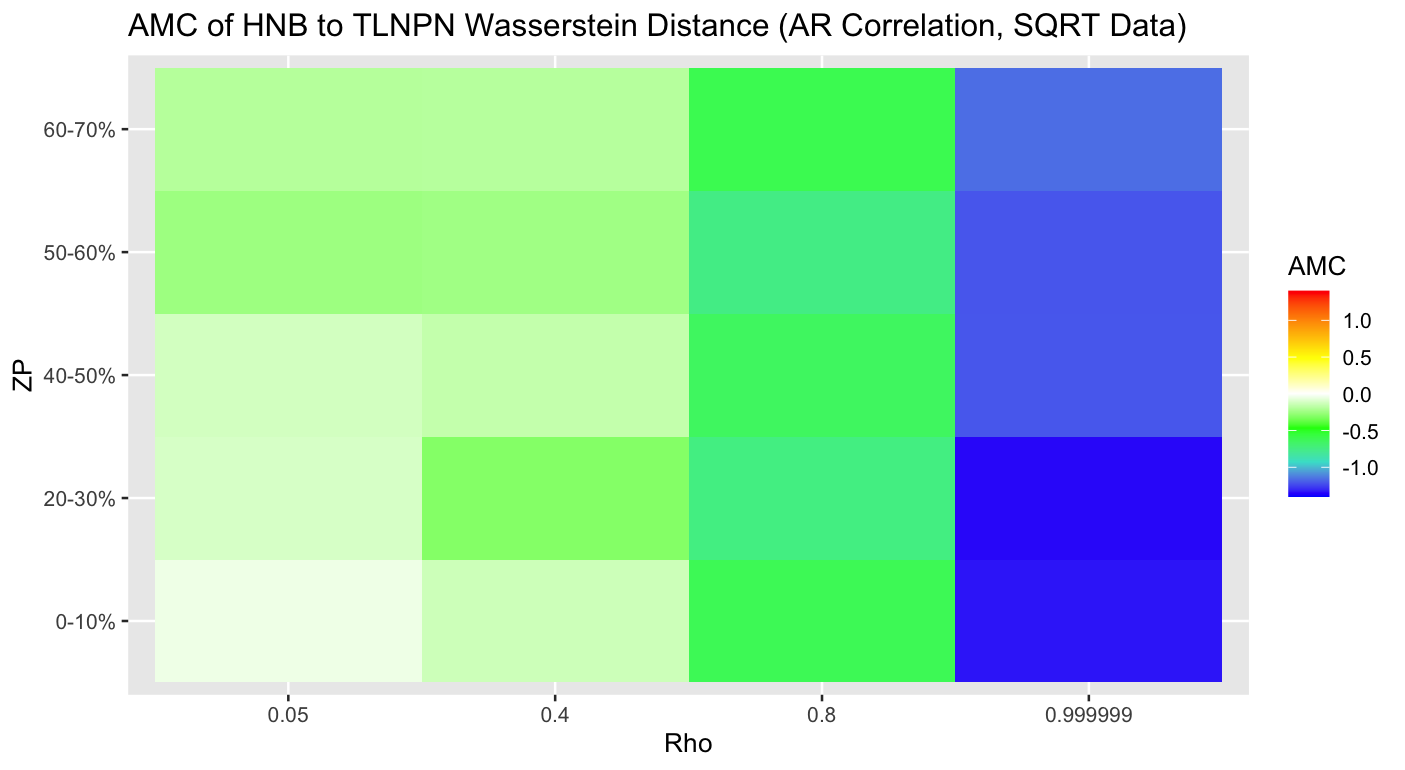}
\caption{In these heat maps, we compare the performance of the TLNPN model to the HNB model where the population data was generated from the TLNPN model where the latent correlation matrix follows an AR correlation structure. On the left, the population TLNPN data was trained on the untransformed data, and on the right, the population TLNPN data was trained on the square root of the QMP data. We see in both heat maps, the TLNPN model universally outperforms the HNB model, and the TLNPN model performs best when $\rho=0.999999$ (i.e., when there is strong dependence among the variables). Additionally, we see that the zero-proportion of the variables does not have a reliable effect on relative model performance.}
\label{fg_HeatAMC_ZP_ARod}
\end{figure}

In \textbf{Figure \ref{fg_HeatAMC_ZP_ARod}}, we present our results for the AR correlation structure. Here, our results are quite similar: the higher $\rho$ is (meaning the higher the correlation between the latent Gaussian variables is), then the better the TLNPN model performs relative to the HNB model. Again, we see that zero-proportion does not impact relative model fit.

\subsection*{\textit{Real Data Analysis: Quantitative Microbiome Profiling Data}}

\begin{figure}[hbt!]
\centering
\includegraphics[width = 0.48\textwidth]{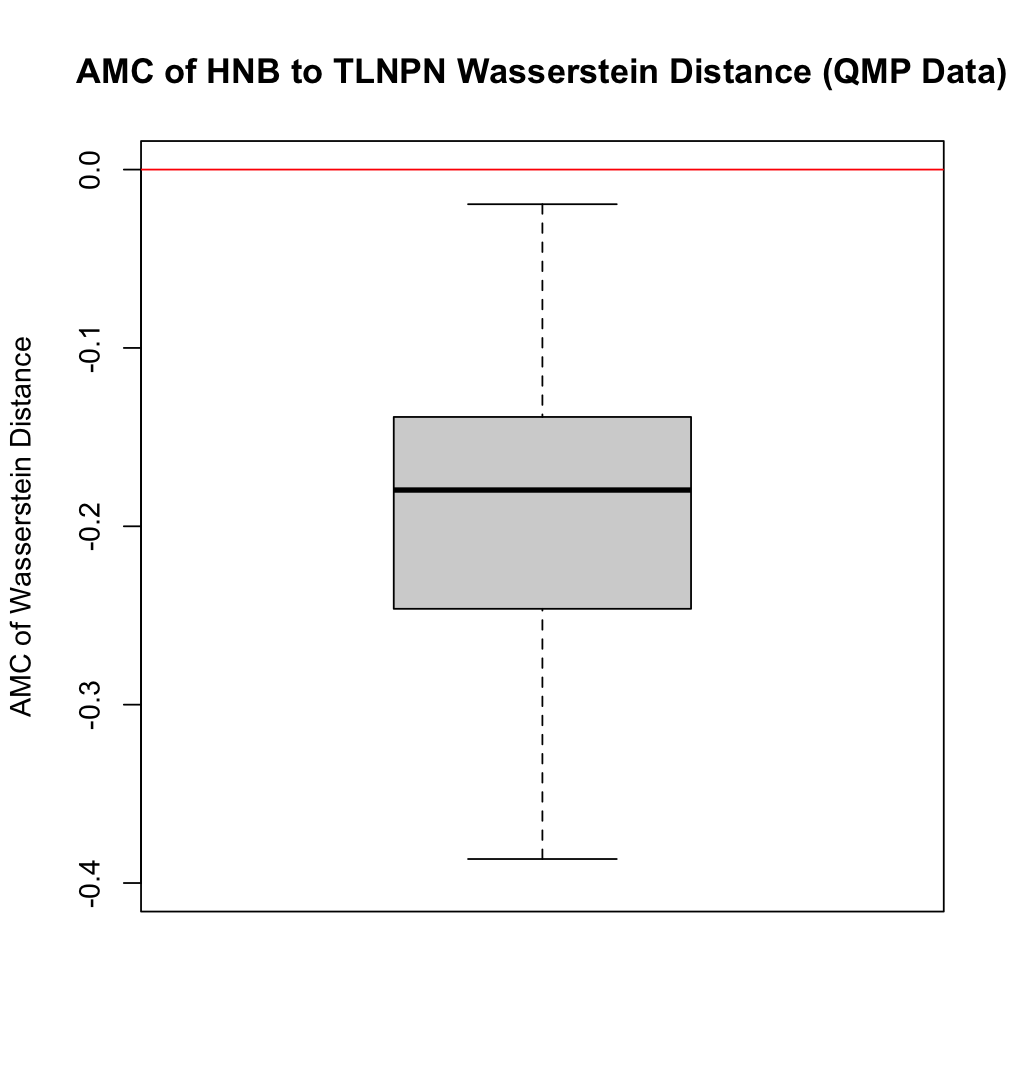}\includegraphics[width = 0.5\textwidth]{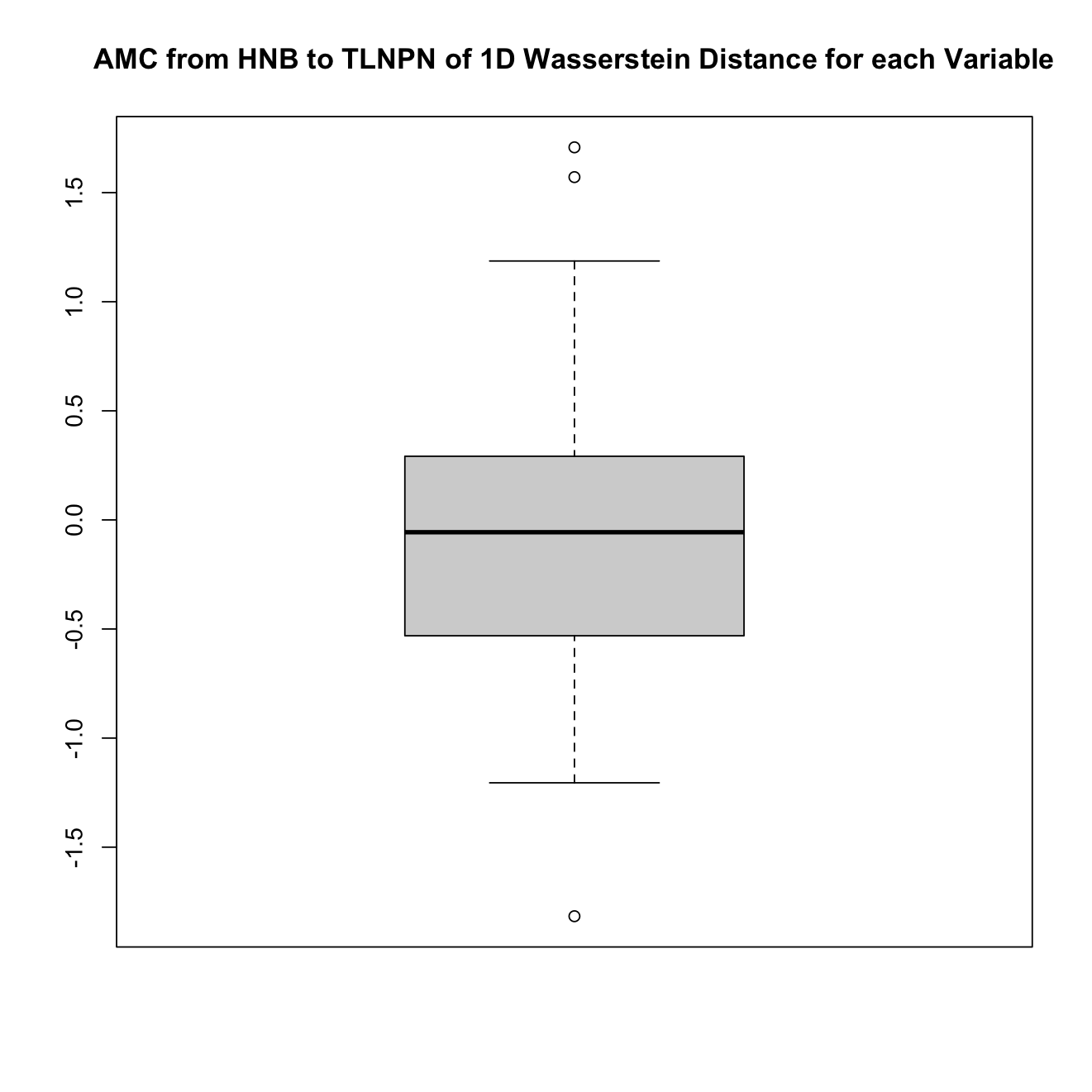}
\caption{Here, we display the result of our QMP data analysis. The first box plot shows the relative performance of the TLNPN model to the HNB model, in which we see that it outperformed the HNB model in all 50 splits. We see in the second box plot that the TLNPN model and the HNB model performed similarly, on average in modeling the marginal distributions. Therefore, the TLNPN model performed better modeling the joint distribution as it could model dependence.}
\label{fg_BoxAMC_QMPData}
\end{figure}

As an example of real world zero-inflated data, we use Quantitative Microbiome Profiling (QMP) \citep{vandeputte2017quantitative}. This data measures the number of 101 different genera of gut bacteria in 135 people (29 with Crohn's Disease and 106 controls). 
We found that a limitation of fitting the HNB model was the most popular R function for fitting the HNB model (from the pscl package) was limited to integers less than or equal to $2^{31} -1$, so we had to rescale the data by taking the power of 
$0.851$ of each data point and then rounding, which makes the maximum value of the data $2^{31} -1$.
The first, second, third, and fourth quartiles of the zero-proportion of the $101$ variables are 3.7\%, 28.9\%, 57.8\%, and 79.3\% respectively, and the data displays a high amount of skewness. The result of the analysis is displayed in \textbf{Figure \ref{fg_BoxAMC_QMPData}}.

We found that the TLNPN model outperformed the HNB model in terms of $p$-dimensional Wasserstein distance in every replication. We also see in \textbf{Figure \ref{fg_BoxAMC_QMPData}} a box plot of the AMC of the one-dimensional Wasserstein distances of the each of the 101 variables from the HNB model to the TLNPN model.

\subsection*{\textit{Real Data Analysis: Single Cell RNA Sequencing Data}}

For our second real data analysis, we use single-cell RNA sequencing data 
from the lymphoblastoid cell line; the original data can be found on the 10x Genomics Datasets website (\textit{https://www.10xgenomics.com}). 
This dataset measures $p=329$ genes from $n=265$ cells. The first, second, third, and fourth quantiles of the zero-proportion of the variables in the RNA dataset are 41.1\%, 65.7\%, 81.1\%, and 89.8\% respectively. In this case, we did not have to transform the data as all data points were already well below $2^{31}$ and $95\%$ are below 10. 
\begin{figure}[hbt!]
\centering
\includegraphics[width = .48\textwidth]{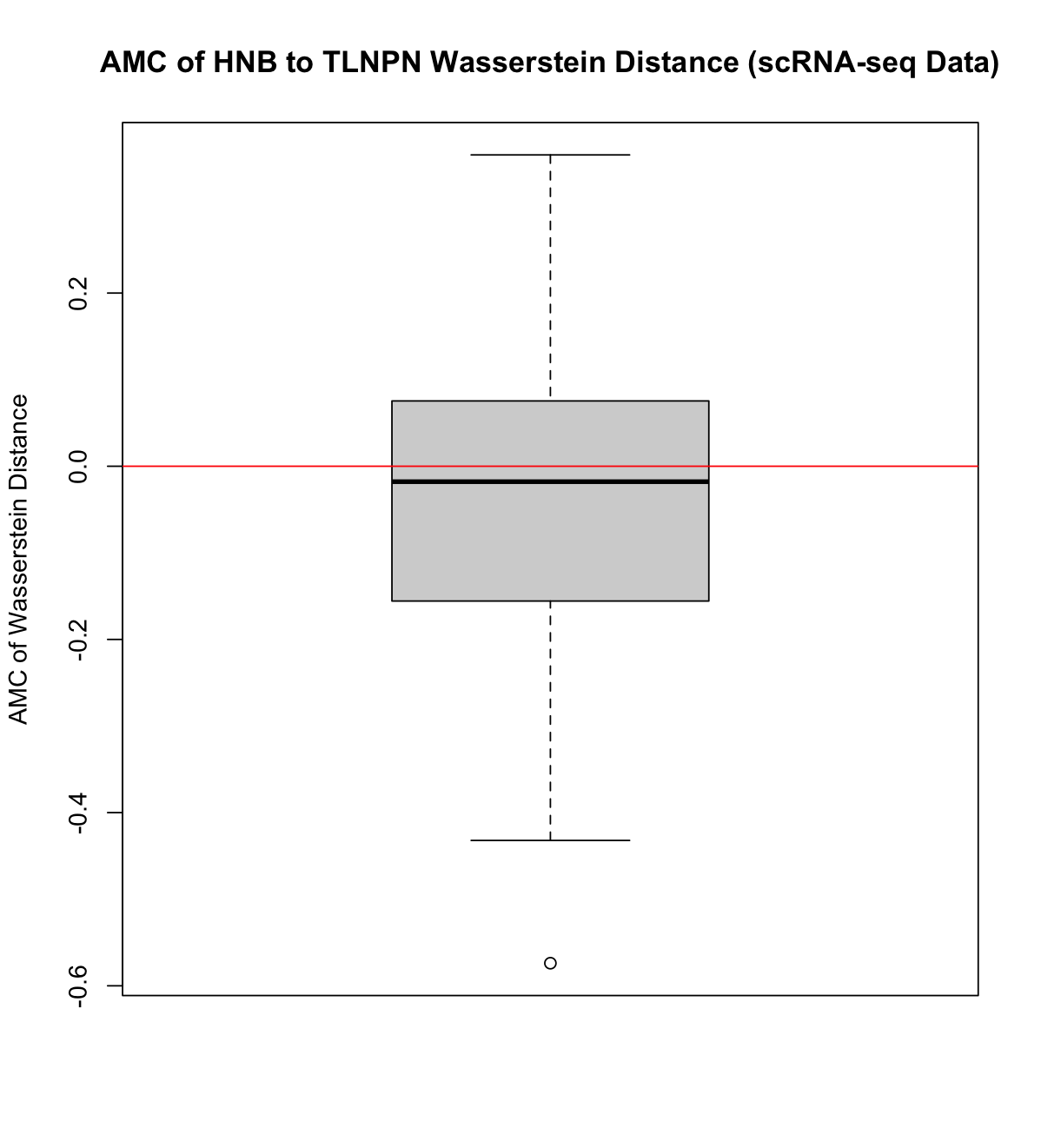}\includegraphics[width = 0.5\textwidth]{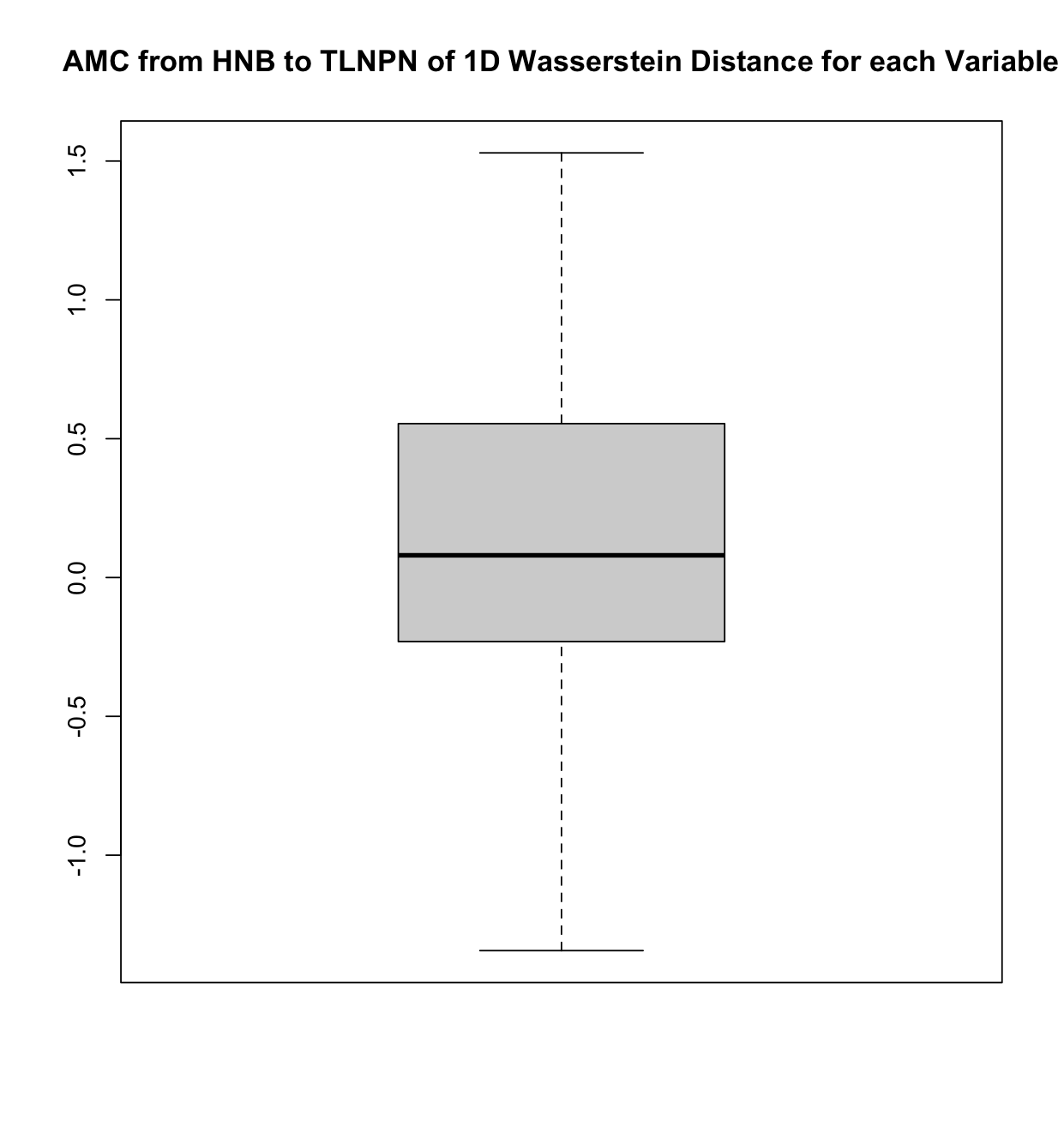}
\caption{Here, we present the result of our gene sequencing data analysis. On the left, we compare the performance of the TLNPN model to the HNB model under the 50 random splits of the gene sequencing data. We see that that the models perform nearly the same on average. The second box plot displays how the models compared modeling the variables on a marginal level where they performed similarly.}
\label{fg_BoxAMC_ATACData}
\end{figure}
However, we found in \textbf{Figure \ref{fg_BoxAMC_ATACData}} that the TLNPN model and HNB model performed similarly over the 50 replications for the $p$-dimensional Wasserstein distance. Furthermore, we found that the HNB and TLNPN models performed similarly modeling the marginal distributions of the $p$ variables.

\section*{DISCUSSION}

\subsection*{\textit{Simulation Studies}}

In simulation setting 1, we found that when $\beta_1=\gamma_1=2$, the ZINB and HNB were vulnerable to model misspecification as seen in \textbf{Figure \ref{fg_ZINBvHNBcv}}. We conclude that this trend is due to the fact that the ZINB model will predict sampling zeros given a low covariate whereas the HNB model would never predict a sampling zero, causing the difference in AIC.
Furthermore, we also observed that in cases of zero-deflation, the HNB model far outperformed the ZINB model, which is displayed in \textbf{Figure \ref{fg_ZeroDeflationZINB.png}}. However, the ZINB model seemed robust to moderate levels of zero-deflation.
This robustness of the ZINB model to moderate zero deflation arises from its ability to adjust parameters such as the dispersion parameter, compensating for a lower-than-expected proportion of zeros. However, as the zero-deflation intensifies, the difference in AIC begins to grow since the ZINB model cannot predict zeros at a probability below that of a standard negative binomial distribution. 

The first main result of our second simulation setting is displayed in \textbf{Figure \ref{fg_HeatAMC_AR_Beta1}} where we see that when the HNB model is fitted without covariates, the TLNPN model performs best when $\beta_1=2$ and $\rho=0.9$ under the AR correlation structure.
We attribute this to the ability of the TLNPN model to account for the correlation between the latent Gaussian variables, which can be influenced through the correlation between the covariates. However, for the correlation between the covariates to have an impact on the latent correlation, the $\beta_1$ parameter, which controls the impact the covariates have on the mean, has to be nonzero. 
We see that regardless of $\rho$, when $\beta_1  = 0$, the models perform nearly identically because the impact of the correlation between the covariates has no bearing on the correlation of the latent variable since the covariates have no effect on the mean. 
Similarly, when $\beta_1 = 2$ and $\rho=0.01$, we see that the models perform nearly the same, because there is a lack of dependence among the covariates and thus, the variables.
Therefore, under the AR correlation structure, when both $\rho$ and $\beta_1$ are large,  we see the best relative performance of the TLNPN model because the variables are more highly dependent on each other since the covariates are highly correlated and the covariates have a large impact on $\mu$.

In \textbf{Figure \ref{fg_HeatAMC_AR_Beta1}}, we also observe that when the HNB model is fitted with covariates, it outperforms the TLNLPN model most when both $\beta_1$ and $\rho$ are high. We conducted a follow-up simulation study to investigate this trend; the results of with are summarized in \textbf{Figure \ref{fg_BoxCord}} and \textbf{Figure \ref{fg_Box1DwdistHNBvTLNPNB2}}. We see in these figures that the TLNPN underestimates the correlation among the HNB variables and performs worse than the HNB model fitted with covariates on the marginal level. 
We attribute this trend to the TLNPN model estimating the latent correlation between the latent Gaussian variables through a formula that utilizes Kendall's $\tau$. In contrast, the HNB model was able to much more accurately capture the correlation between the variables.
We attribute this to the HNB model using the covariates of the test data when simulating data from the fitted model, resulting in more accurate predictions than those of the TLNPN model, particularly for extremely high values. The TLNPN model can only predict values within its training dataset, which makes it vulnerable to modeling datasets with extreme, outlier values, which are much more probable when $\beta_1=2$ as compared to $1$ or $0$. 

In \textbf{Figure \ref{fg_HeatAMC_AR_Gamma1_B0_ncv}}, we see that when $\beta_1=0$, the $\gamma_1$ parameter seems to have no effect on relative model performance.
This is a result of the lack of dependence and extreme values among the variables that results when $\beta_1=0$, so the parameter $\gamma_1$ can only have a limited impact on the dependence and variance of the variables.
However, when $\beta_1=2$, we see a clear pattern emerge both when the HNB model is fitted with covariates and when it is not as shown in \textbf{Figure \ref{fg_HeatAMC_AR_Gamma1_B2_ncv}}. When covariates are not considered when fitting the HNB model, the TLNPN model outperforms the HNB model most when $\rho=0.9$ and $\gamma_1=-0.8$. 
We attribute this trend to the fact that $\beta_1$ is always nonnegative in our simulations, therefore, if an increase in the covariate both increases the mean and decreases the probability of a zero, then the resulting latent correlation, calculated from Kendall's $\tau$ will be much stronger, which the TLNPN model accommodates. 

Despite this, when covariates are considered when fitting the HNB model, the pattern reverses, and the TLNPN model performs worse when $\gamma_1=-0.8$ and $\rho$ is large. We conducted a follow-up simulation study to investigate, and the result is displayed in \textbf{Figure \ref{fg_HistResidualTLNPNvHNB}} where we see that as $\gamma_1$ increases, the residuals between the simulated data and the test data greatly reduces, particularly for the TLNPN data. 
Therefore, there are two trends at work that cause the TLNPN model to perform worse against the HNB model fitted with covariates when $\gamma_1 =-0.8$ as compared to when $\gamma_1 = 0.8$ and $\rho$ is high. 
One, when $\gamma_1 = -0.8$, the probability of a structural zero decreases as the covariate, and therefore the mean of the distribution, increases. This increases the correlation between the zero-inflated variables given $\rho$ is high, and the HNB model more accurately describes the correlation structure between the zero-inflated variables than the TLNPN model, which tends to underestimate the correlation between the variables. Two, as the mean of the HNB distribution increases, the variance increases as well, which will increase the Wasserstein distance between test and simulated data. However, the HNB is better equipped to predict higher values compared to the TLNPN model because that model uses the same covariates as the test data. We see that when $\gamma_1 = 0.8$, there is an improvement in the relative performance of the TLNPN model against the HNB model fitted with covariates modeling the marginal distribution compared to when $\gamma_1 = -0.8$ as seen in \textbf{Figure \ref{fg_Box1DwdistHNBvTLNPNB2}}.
Therefore, the TLNPN model performs relatively better when $\gamma_1 = 0.8$ because it makes the occurrence of an extremely high data point less probable, which we see in \textbf{Figure \ref{fg_HistResidualTLNPNvHNB}} where as $\gamma_1$ increases, the size of the residuals between the marginal distributions decreases dramatically. 
We also found that the level of zero-inflation or deflation had no effect on the relative model performance.
We conclude that this results from the ability of both models to account for zero-inflation and zero-deflation.

We also performed the simulation study where the covariates were generated from a GD correlation structure.
Under the GD correlation structure, we see results investigating the interaction between $\rho$ and $\beta_1$ in \textbf{Figure \ref{fg_HeatAMC_GD_Beta1_ncv}} where when the HNB model is fitted without covariates, the TLNPN model outperforms the HNB model when $\rho=0.01$ and $\beta_1=2$. We conclude that this results from the dependence among the HNB variables that results when the covariates are highly correlated and have a large impact on $\mu_{ij}$, which the TLNPN model accounts for, but the HNB does not.
We see that when the HNB model is fitted with covariates, the HNB model outperforms the TLNPN model most when $\beta_1=2$ and $\rho=0.01$ since it more accurately describes the dependency structure and can better predict large values. For both scenarios, we see that when $\beta_1 = 0$, the models perform nearly identically as the zero-inflated variables have almost no dependence among each other, and both models fit the marginal distributions similarly well. 

We also investigated the interaction between $\rho$ and $\gamma_1$ under the GD correlation structure, which is presented in \textbf{Figure \ref{fg_HeatAMC_GD_Gamma1}}. 
The interpretation of these results are the same as the interpretation of the results when the covariates are generated from an AR correlation structure; when $\beta_1 = 0$, then $\gamma_1$ has very little impact on the dependence structure of the zero-inflated variables, so it doesn't have an impact on the relative performance between the TLNPN and the HNB models regardless of whether the HNB model was fitted with covariates. 
However, when $\beta_1 > 0$ and $\gamma_1 < 0$, then the correlation between the zero-inflated variables will strengthen since the higher the mean of the distribution is, the lower the probability of a structural zero. When the HNB model is fitted without covariates, the TLNPN model outperforms the HNB model the most when $\rho = 0.01$ and $\gamma_1 = -0.8$; however, when the HNB model is fitted with covariates, the opposite is true due to the HNB model better describing the dependence structure and marginal distributions of the zero-inflated variables. 

In \textbf{Figure \ref{fg_HeatAMC_ZP_GDod}}, we display results from the third simulation setting where we investigate the effect of zero-proportion, $\rho$ (under the GD correlation structure) on relative model performance, and the square-root data transformation. We see that in the square-root transformation, the TLNPN model outperforms the HNB model most when $\rho=0.05$.
We can attribute this to the higher correlation between the zero-inflated variables, which the TLNPN model is able to account for as opposed to the HNB model. However, it still seems that zero-proportion doesn't have an effect on the relative model performance. We attribute this to the fact that both models can handle zero-inflation or deflation.
We display our results for the AR correlation structure in \textbf{Figure \ref{fg_HeatAMC_ZP_ARod}}, and see in both the untransformed CDF and square-root transformation CDF, the TLNPN model improves its performance against the HNB model as $\rho$ increases. 
We again conclude that this is due to the TLNPN model's ability to model dependence among the zero-inflated variables. 

\subsection*{\textit{Real Data Analyses}}

Our first real data analysis compared the performance of the TLNPN model against the HNB model using the QMP dataset; the results are displayed in \textbf{Figure \ref{fg_BoxAMC_QMPData}} where we see that the TLNPN model outperformed the HNB model with regards to $p$-dimensional Wasserstein distance under every random split but performed similarly, on average, on a marginal level.
We conclude that the difference between the Wasserstein distances was a result of the TLNPN model accounting for dependence among the variables whereas the HNB model does not model dependence among the variables.
We see that the two models perform, on average, about the same, which rules out the explanation that the TLNPN model outperformed the HNB model due to marginal fit. 

\begin{figure}[hbt!]
\centering
\includegraphics[width = 0.95\textwidth]{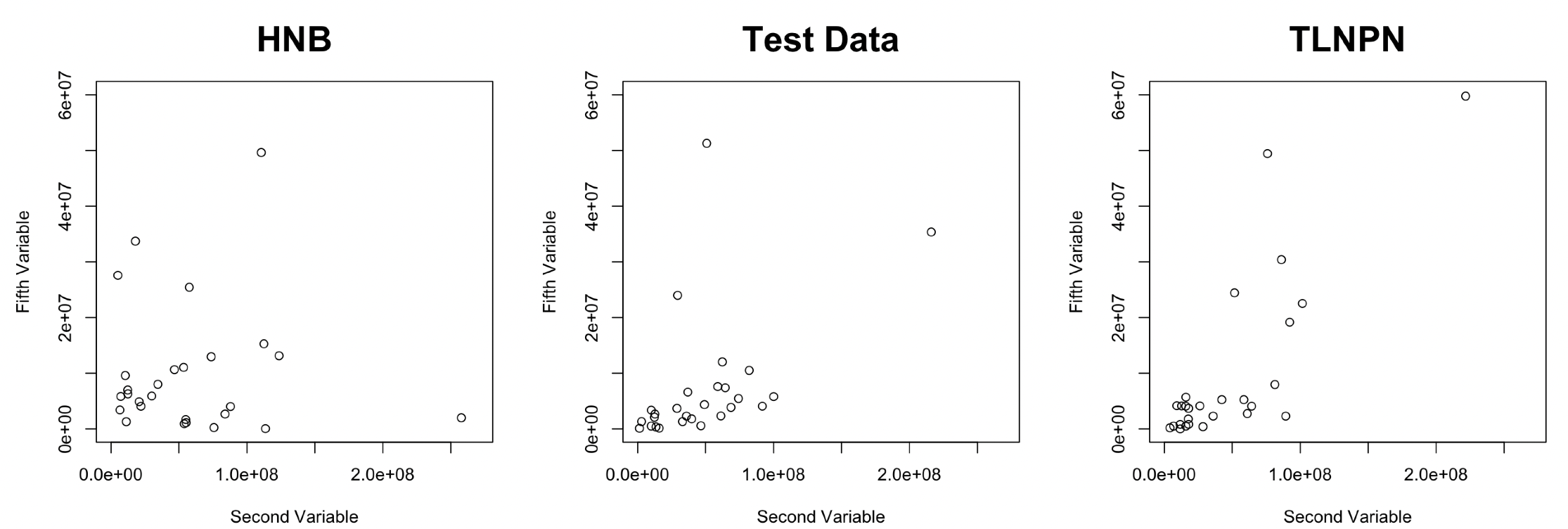}
\caption{Here, we compare the joint distributions of second and fifth variables of the HNB, TLNPN, and QMP test data to show how the TLNPN model is able to model the dependence of the test data that the HNB model cannot.}
\label{fg_Scatter_QMP_HNBvTest}
\end{figure}

To further emphasize this difference, we use the example of the second and fifth variables, which were significantly correlated with each other, and compare how the HNB and TLNPN models modeled their joint distribution as compared with the test data. We see in \textbf{Figure \ref{fg_Scatter_QMP_HNBvTest}} that there is a dependence between the variables in the test data, which the TLNPN data captures, but the HNB data does not, leading to a higher Wasserstein distance for the HNB simulated data. 

In our second real-data analysis, we compared the TLNPN model to the HNB model using single cell RNA sequencing data; the results of which are displayed in \textbf{Figure \ref{fg_BoxAMC_ATACData}}. We see that the models perform similarly on a multivariate and marginal level.
Based on these results, we conjecture that the characteristics of this dataset, the small scale and a lack of highly correlated variables, resulted in the similar performance of the HNB and TLNPN models. The TLNPN model is more robust to extreme values than the HNB model fitted without covariates, but this dataset had very little skewness and variance in comparison to the QMP dataset, which contributed to the TLNPN model performing similarly to the HNB model.

\section*{CONCLUSION}

In this report, we sought to compare models for zero-inflated data through both simulation and real data studies that mimicked and used modern biomedical data. Zero-inflated and hurdles models have been popularly used in this field and we sought to compare them with the newly introduced truncated latent Gaussian copula model. We found in the simulation studies and real data analyses that the main considerations for deciding to fit either the TLNPN or the HNB models were access to covariates, variance of the data, and dependence among the variables. 

The obvious advantage of using the TLNPN model is that it can account for dependence among variables without having access to the covariates in contrast to the HNB model. However, the TLNPN requires a large amount of training data to accurately estimate the correlation of the Gaussian-level variables, and furthermore, in cases of strong dependence among the variables, the TLNPN model tends to underestimate the correlation between the zero-inflated variables when the true model is a HNB model. Furthermore, when the HNB model has access to the covariates, it tends to model the dependence structure between the variables much more accurately. Nevertheless, when no covariate is available, the TLNPN model typically outperforms the HNB model in fitting multivariate distributions of highly dependent zero-inflated variables.

Another drawback of the TLNPN model is that when the true population is HNB with a high $\beta_1$ parameter, the TLNPN model struggles to model large, outlier values compared to the HNB fitted with covariates. The TLNPN model will never predict a data point outside of its original training data because of its use of the empirical CDF, so the training data must be similar to the testing data for it to perform well. 
However, on a marginal level, the HNB model itself can be vulnerable to overdispersed and highly skewed data, which the TLNPN model is better at fitting marginally. Furthermore, a computational limitation of the HNB model is that the main function used to fit the data to a HNB model can only handle integer values below $2^{31}$, so datasets with larger values need to be rescaled until this issue is addressed as many modern biomedical datasets contain values greater than $2^{31}$.

Future research could investigate how the TLNPN model performs against other models, particularly zero-inflated Poisson and hurdle Poisson models with an overdispersion parameter. Furthermore, investigation of incorporating covariates into the TLNPN model will be an interesting research direction to pursue.

\section*{ACKNOWLEDGEMENTS}

This research was supported by the 2024 Mathematics Research Experiences for Undergraduates (REU) program at the University of North Carolina at Charlotte, under NSF-REU grant DMS-2150179.

\clearpage

\section*{APPENDIX}

\section{Figures}
\begin{figure}[!htb]
	\centering
    \includegraphics[width=0.49\textwidth]{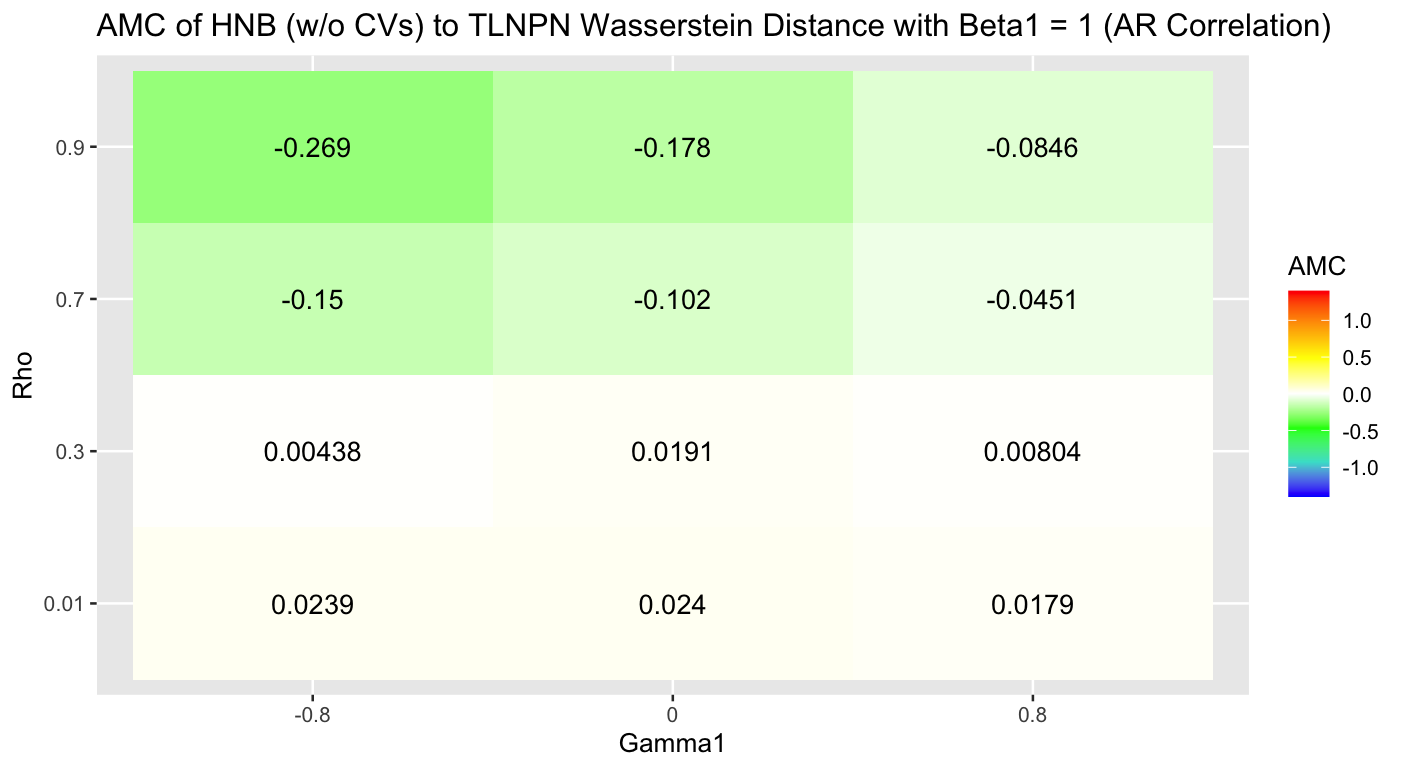}
    \includegraphics[width=0.49\textwidth]{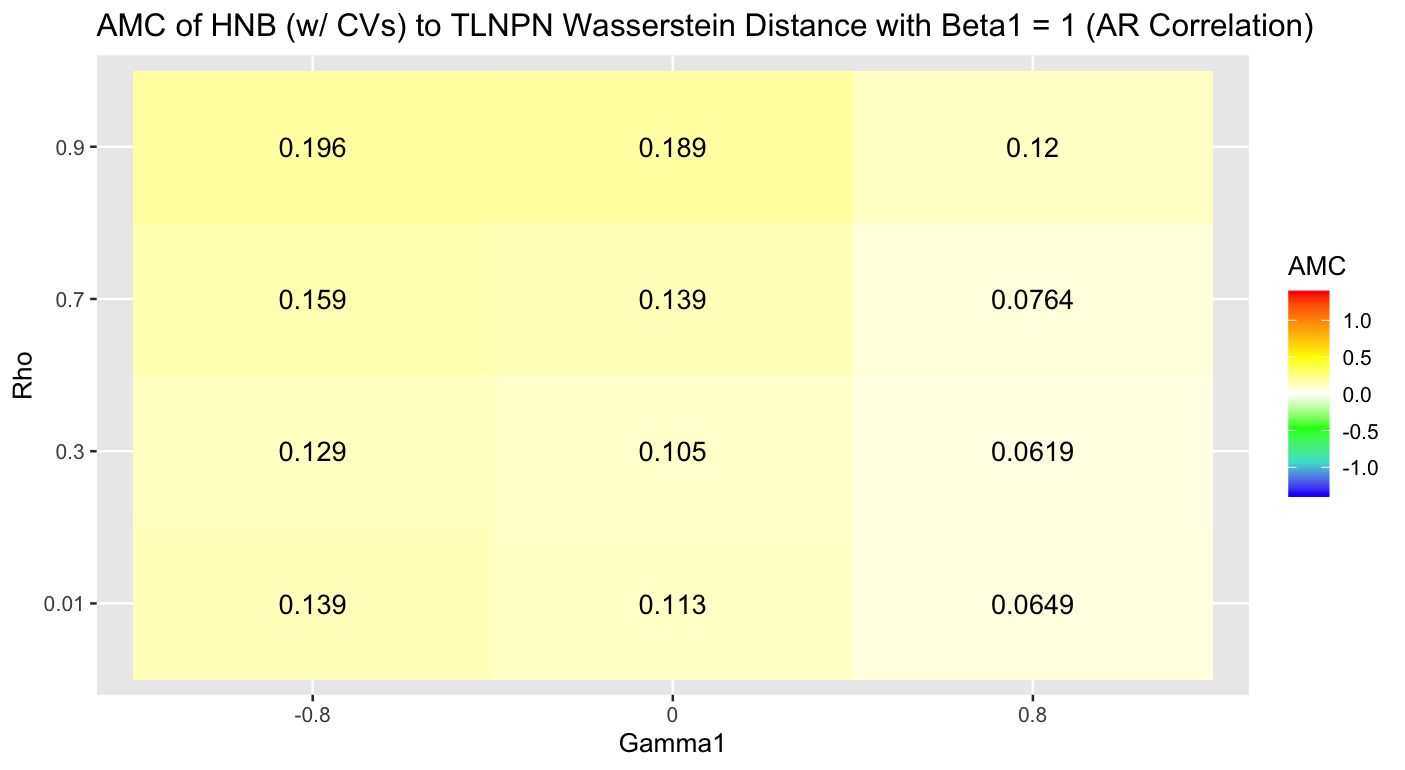}	
    \caption{ Heatmap of AMC of Wasserstein distances with and without covariates.}
\end{figure}

\begin{figure}
\centering
\includegraphics[width = 0.6\textwidth]{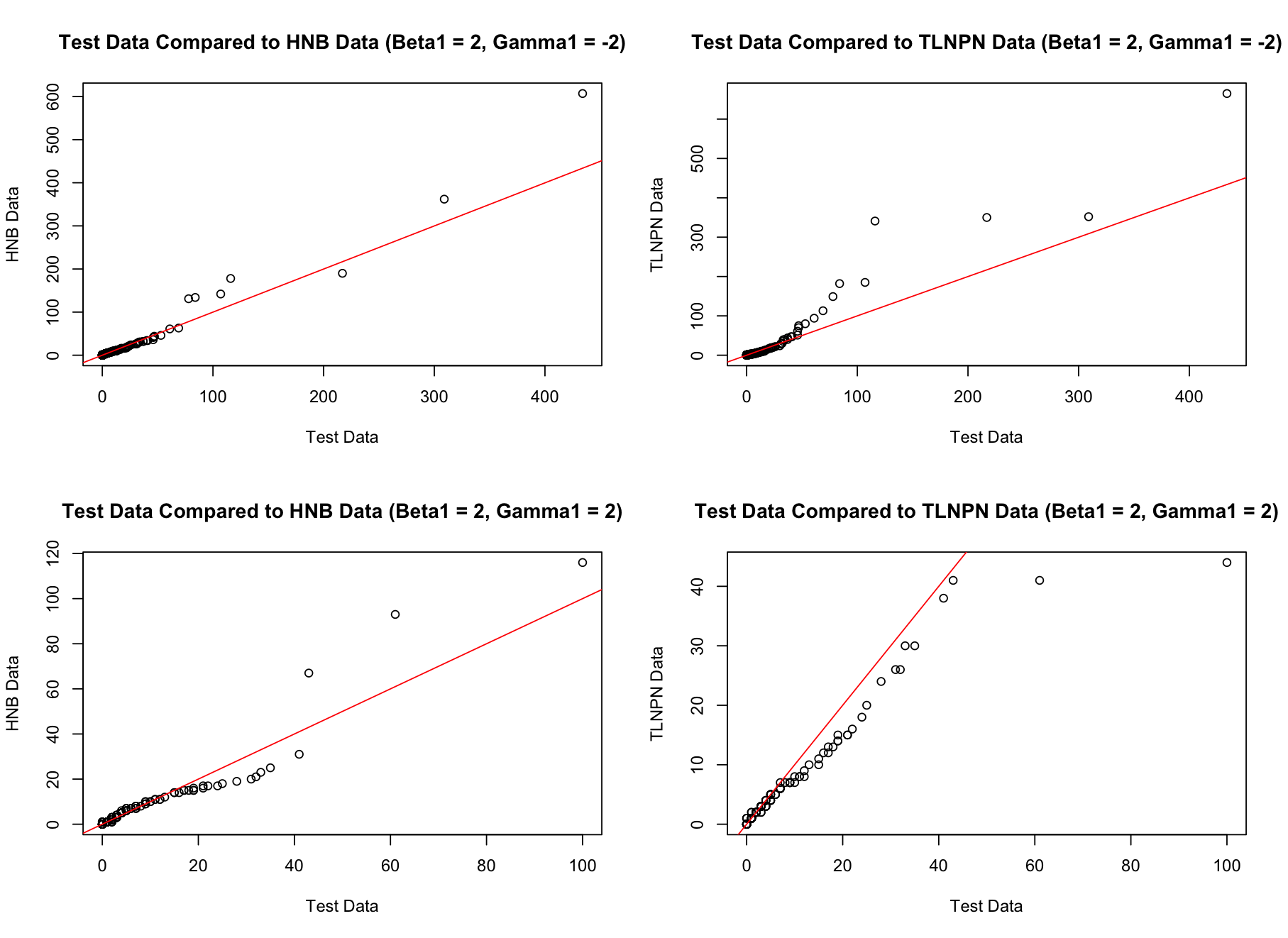}
\caption{This shows the scatterplots of the Test vs. the TLNPN and HNB simulated data under varying $\gamma_1$ conditions. }
\end{figure}

\begin{figure}
\centering
\includegraphics[width=0.6\textwidth]{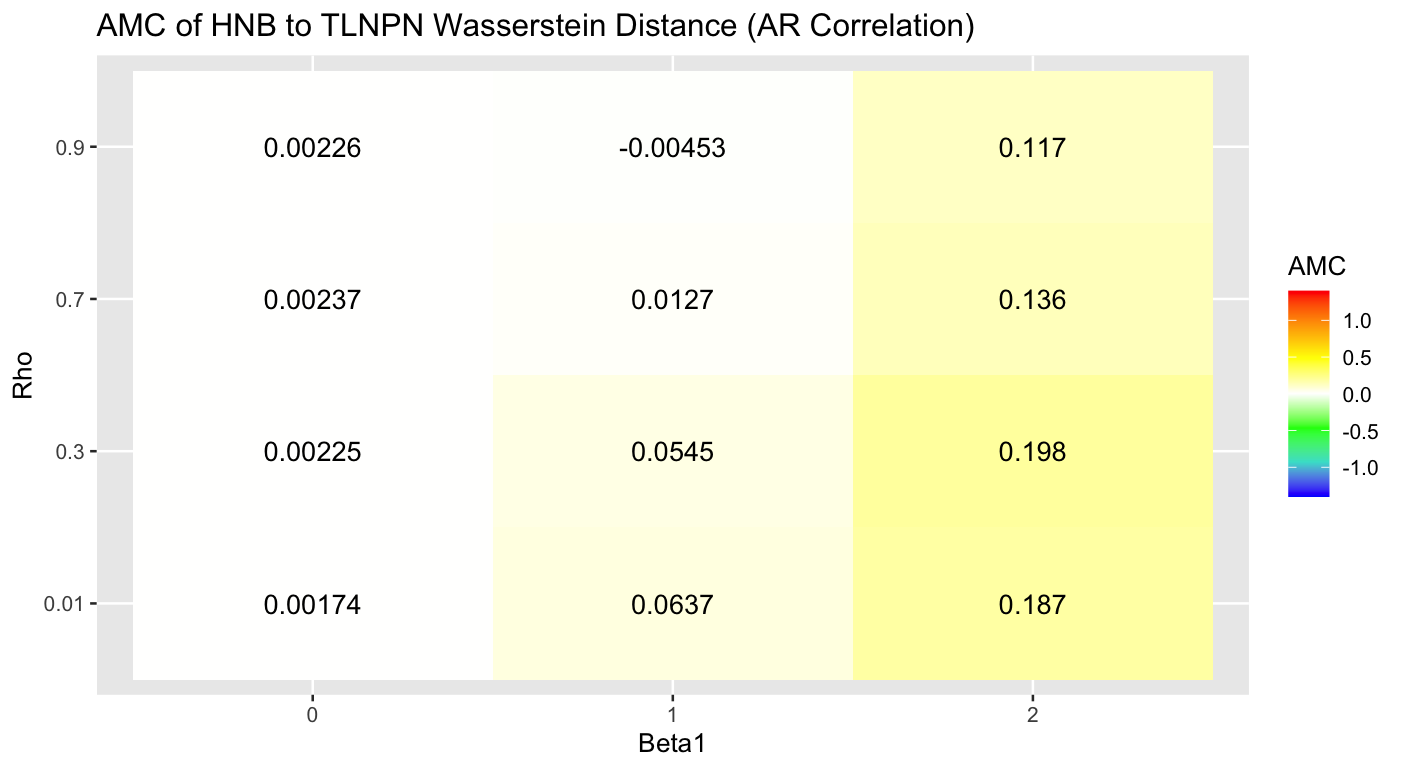}
\caption{Heatmap of the arithmetic mean changes of Wasserstein distances under HNB population with AR covariance matrix.}
\end{figure}

\begin{figure}
\centering
\includegraphics[width = 0.6\textwidth]{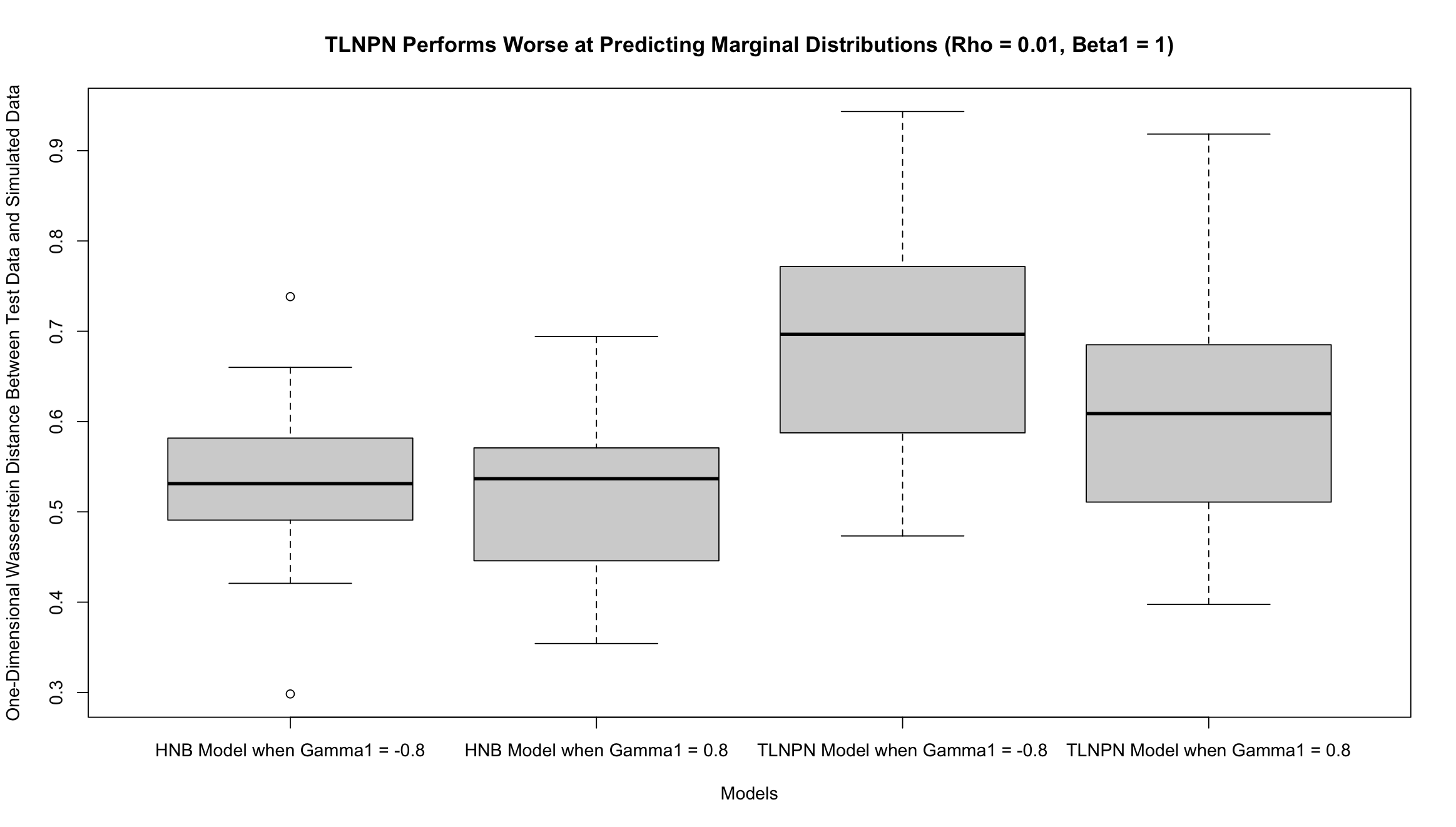}
\caption{This shows the difference in the TLNPN model predicting marginal distributions as compared to the HNB model under varying $\gamma_1$ conditions.}
\end{figure}

\begin{figure}
\centering
\includegraphics[width=0.6\textwidth]{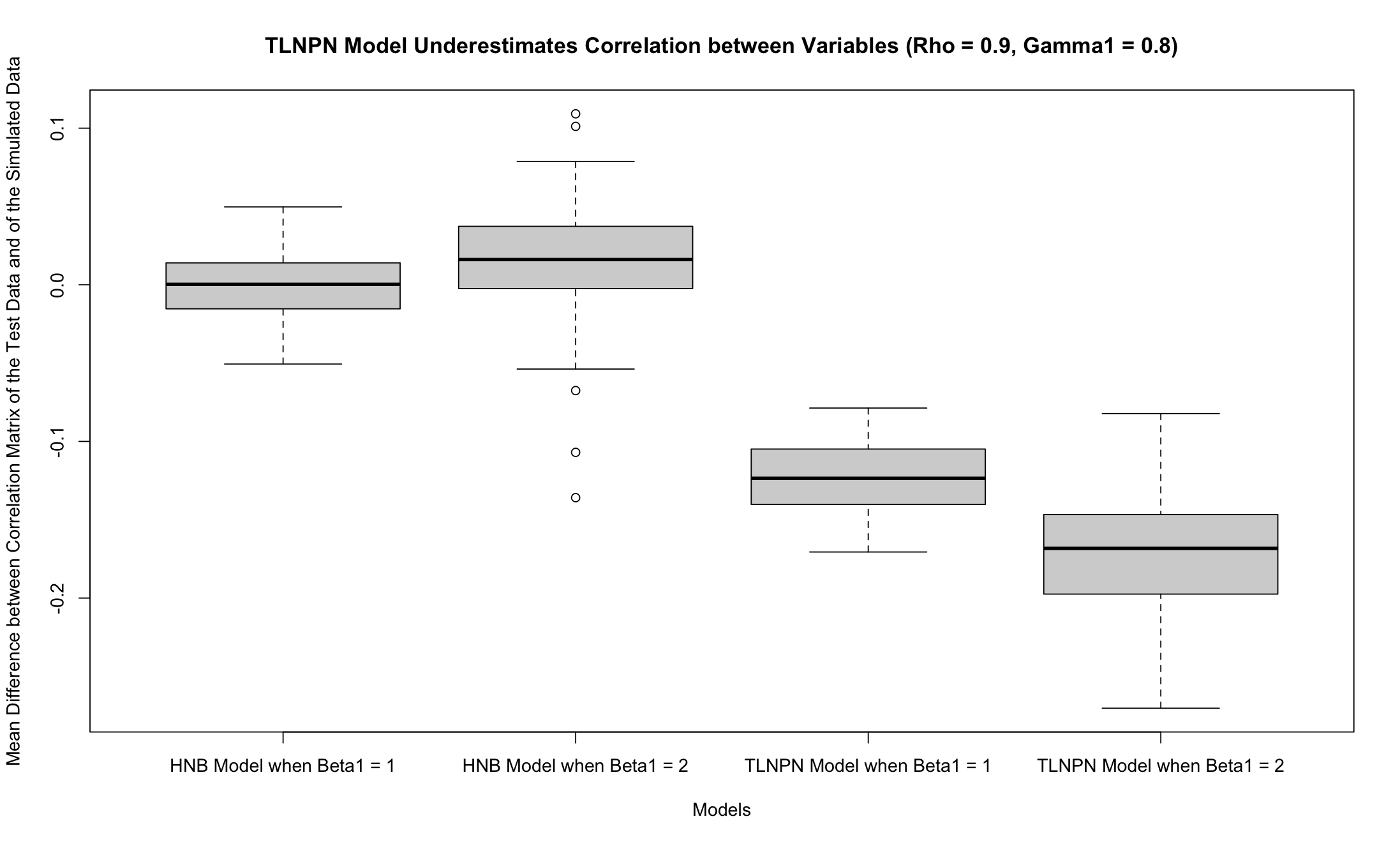}
\caption{This shows the difference in the correlation matrices under varying $\beta_1$ conditions and models.}
\end{figure}

\clearpage
\bibliographystyle{plainnat}
\bibliography{ref}

\begin{thebibliography}{15}
\providecommand{\natexlab}[1]{#1}
\providecommand{\url}[1]{\texttt{#1}}
\expandafter\ifx\csname urlstyle\endcsname\relax
  \providecommand{\doi}[1]{doi: #1}\else
  \providecommand{\doi}{doi: \begingroup \urlstyle{rm}\Url}\fi

\bibitem[Campbell(2021)]{https://doi.org/10.1111/2041-210X.13559}
Harlan Campbell.
\newblock The consequences of checking for zero-inflation and overdispersion in the analysis of count data.
\newblock \emph{Methods in Ecology and Evolution}, 12\penalty0 (4):\penalty0 665--680, 2021.
\newblock \doi{https://doi.org/10.1111/2041-210X.13559}.
\newblock URL \url{https://besjournals.onlinelibrary.wiley.com/doi/abs/10.1111/2041-210X.13559}.

\bibitem[Chung et~al.(2022)Chung, Ni, and Gaynanova]{chung2022sparse}
Hee~Cheol Chung, Yang Ni, and Irina Gaynanova.
\newblock Sparse semiparametric discriminant analysis for high-dimensional zero-inflated data.
\newblock \emph{arXiv preprint arXiv:2208.03734}, 2022.

\bibitem[Dong et~al.(2014)Dong, Clarke, Yan, Khattak, and Huang]{dong2014multivariate}
Chunjiao Dong, David~B Clarke, Xuedong Yan, Asad Khattak, and Baoshan Huang.
\newblock Multivariate random-parameters zero-inflated negative binomial regression model: An application to estimate crash frequencies at intersections.
\newblock \emph{Accident Analysis \& Prevention}, 70:\penalty0 320--329, 2014.

\bibitem[Ehsan~Saffari et~al.(2012)Ehsan~Saffari, Adnan, and Greene]{ehsan2012hurdle}
Seyed Ehsan~Saffari, Robiah Adnan, and William Greene.
\newblock Hurdle negative binomial regression model with right censored count data.
\newblock \emph{SORT: statistics and operations research transactions}, 36\penalty0 (2):\penalty0 0181--194, 2012.

\bibitem[Fan et~al.(2017)Fan, Liu, Ning, and Zou]{fan2017high}
Jianqing Fan, Han Liu, Yang Ning, and Hui Zou.
\newblock High dimensional semiparametric latent graphical model for mixed data.
\newblock \emph{Journal of the Royal Statistical Society Series B: Statistical Methodology}, 79\penalty0 (2):\penalty0 405--421, 2017.

\bibitem[Feng(2021)]{feng2021comparison}
Cindy~Xin Feng.
\newblock A comparison of zero-inflated and hurdle models for modeling zero-inflated count data.
\newblock \emph{Journal of statistical distributions and applications}, 8\penalty0 (1):\penalty0 8, 2021.

\bibitem[Freckleton(2009)]{https://doi.org/10.1111/j.1420-9101.2009.01757.x}
R.~P. Freckleton.
\newblock The seven deadly sins of comparative analysis.
\newblock \emph{Journal of Evolutionary Biology}, 22\penalty0 (7):\penalty0 1367--1375, 2009.
\newblock \doi{https://doi.org/10.1111/j.1420-9101.2009.01757.x}.
\newblock URL \url{https://onlinelibrary.wiley.com/doi/abs/10.1111/j.1420-9101.2009.01757.x}.

\bibitem[Hua et~al.(2014)Hua, Wan, Wenjuan, and Crits-Christoph]{hua2014structural}
HE~Hua, TANG Wan, WANG Wenjuan, and Paul Crits-Christoph.
\newblock Structural zeroes and zero-inflated models.
\newblock \emph{Shanghai archives of psychiatry}, 26\penalty0 (4):\penalty0 236, 2014.

\bibitem[Li(2015)]{li2015microbiome}
Hongzhe Li.
\newblock Microbiome, metagenomics, and high-dimensional compositional data analysis.
\newblock \emph{Annual Review of Statistics and Its Application}, 2\penalty0 (1):\penalty0 73--94, 2015.

\bibitem[Liu et~al.(2009)Liu, Lafferty, and Wasserman]{liu2009nonparanormal}
Han Liu, John Lafferty, and Larry Wasserman.
\newblock The nonparanormal: Semiparametric estimation of high dimensional undirected graphs.
\newblock \emph{Journal of Machine Learning Research}, 10\penalty0 (10), 2009.

\bibitem[Panaretos and Zemel(2018)]{PanaretosandZemel2018}
Victor Panaretos and Yoav Zemel.
\newblock Statistical aspects of wasserstein distances.
\newblock \emph{Annual Review of Statistics And Its Application}, 6, 2018.

\bibitem[Peruman-Chaney et~al.(2013)]{Peruman-Chaney2013}
Suzanne Peruman-Chaney et~al.
\newblock Zero-inflated and overdispersed: what's one to do?
\newblock \emph{International Journal of Statistics and Probability}, 2\penalty0 (2):\penalty0 59--67, 2013.

\bibitem[Rose et~al.(2006)Rose, Martin, Wannemuehler, and Plikaytis]{rose2006use}
Charles~E Rose, Stacey~W Martin, Kathleen~A Wannemuehler, and Brian~D Plikaytis.
\newblock On the use of zero-inflated and hurdle models for modeling vaccine adverse event count data.
\newblock \emph{Journal of biopharmaceutical statistics}, 16\penalty0 (4):\penalty0 463--481, 2006.

\bibitem[Vandeputte et~al.(2017)Vandeputte, Kathagen, D’hoe, Vieira-Silva, Valles-Colomer, Sabino, Wang, Tito, De~Commer, Darzi, et~al.]{vandeputte2017quantitative}
Doris Vandeputte, Gunter Kathagen, Kevin D’hoe, Sara Vieira-Silva, Mireia Valles-Colomer, Jo{\~a}o Sabino, Jun Wang, Raul~Y Tito, Lindsey De~Commer, Youssef Darzi, et~al.
\newblock Quantitative microbiome profiling links gut community variation to microbial load.
\newblock \emph{Nature}, 551\penalty0 (7681):\penalty0 507--511, 2017.

\bibitem[Yoon et~al.(2020)Yoon, Carroll, and Gaynanova]{yoon2020sparse}
Grace Yoon, Raymond~J Carroll, and Irina Gaynanova.
\newblock Sparse semiparametric canonical correlation analysis for data of mixed types.
\newblock \emph{Biometrika}, 107\penalty0 (3):\penalty0 609--625, 2020.

\end{thebibliography}

\end{document}